\documentclass[authoryear]{FLO_v1}%

\usepackage{graphicx}
\usepackage{upgreek}
\usepackage{multicol,multirow}
\usepackage{amsmath,amssymb,amsfonts}
\usepackage{mathrsfs}
\usepackage{amsthm}
\usepackage[figuresright]{rotating}
\usepackage[authoryear]{natbib}
\usepackage{ifpdf}
\usepackage[T1]{fontenc}
\usepackage{newtxtext}
\usepackage{newtxmath}
\usepackage{textcomp}
\usepackage{xcolor}
\usepackage[title]{appendix}

\usepackage[colorlinks,allcolors=blue]{hyperref}
\definecolor{jourcolor}{cmyk}{1,0.57,0.01,0.38}
\hypersetup{
    colorlinks,%
    citecolor=jourcolor,%
    filecolor=jourcolor,%
    linkcolor=jourcolor,%
    urlcolor=jourcolor
}

\theoremstyle{definition}

\articletype{RESEARCH ARTICLE}

\DOI{}

\Year{}

\Vol{}

\Price{}

\art-id{}

\Enumber{}

\begin{document}

\title[Virus transmission by aerosol transport during short conversations]{Virus transmission by aerosol transport during short conversations}



\author[]{Rohit Singhal$^{1,\dagger}$}
\author[]{S. Ravichandran$^{2,\ast,\dagger}$}
\author[3]{Rama Govindarajan}
\author[]{Sourabh S. Diwan$^{1,\ast,\dagger}$}


\address[1]{Department of Aerospace Engineering, Indian Institute of Science, Bengaluru, KA 560012, India}
\address[2]{Nordic Institute for Theoretical Physics, KTH Royal Institute of Technology and Stockholm University, Stockholm, 10691-SE, Sweden}
\address[3]{International Centre for Theoretical Sciences, Tata Institute of Fundamental Research, Bengaluru, KA 560089, India}

\corres{*}{Corresponding author. E-mail:
\emaillink{ravichandran@su.se, sdiwan@iisc.ac.in}}
\address[$\dagger$]{These authors contributed equally to the work.}



\abstract
{Pathogens like the SARS-CoV-2 are transmitted not only through violent expiratory events like coughing, but also through routine activities like breathing/speaking/singing. We perform direct numerical simulations of the turbulent transport of potentially infectious aerosols in short conversations. It is shown that a two-way conversation significantly reduces the aerosol exposure compared to a relative monologue by one person and relative silence of the other. This is because the interaction of the jets ejected from the mouth of each speaker produce a “canceling” effect. Unequal conversation is shown to significantly increase the risk of infection to the person who talks less. Interestingly, a small height difference is worse for infection spread, due to reduced interference between the two speech jets, than two faces at the same level! For small axial separation, speech jets show large oscillations and reach the other person intermittently. We suggest a range of lateral separations between two people to minimize transmission risk. A realistic estimate of the infection probability is provided by including exposure through eyes and mouth, in addition to the more common method of using inhaled virions alone. We expect that our results will provide useful inputs to epidemiological models and to disease management.
}
\maketitle
\begin{boxtext}
\textbf{\mathversion{bold}Impact Statement}
Asymptomatic transmission, through activities like speaking, is believed to be an important route of COVID-19 spread across the world. The evolving turbulent jet that emerges from the mouth of a speaker can carry aerosols to large distances. While direct numerical simulations of turbulent speech flow are becoming common, very little is understood about two-way conversations.  We simulate these to obtain spatio-temporal distributions of aerosols and calculate infection probabilities. We show that the interaction between the two speech jets plays a key role in determining this probability. The risk of infection is reduced considerably during a “dialogue” as compared to a “monologue” and a small height difference is found to be more dangerous than the two speakers of equal heights. The present results can inform public health guidelines for minimizing risk of transmission, such as introducing a lateral separation between conversing people.
\end{boxtext}

\section{Introduction}

SARS-CoV-2, responsible for the COVID-19 pandemic, is known to be transmitted through more than one mode. Around the beginning of the pandemic (in early 2020), the primary mode of infection was believed to be through droplet transmission by an infected symptomatic person, through violent expiratory events like coughing and sneezing. Based on this, WHO recommended maintaining a physical distance of $1$m between people to minimize the spread of infection, whereas the recommendation of CDC-USA is a separation of 6ft.
By mid-2020, it became clear that asymptomatic transmission of the virus was equally likely, through the virus-laden droplets released by an infected person while talking, singing, breathing etc. (Asadi et al., \hyperlink{bib2}{2020}; Morawska {\&} Cao, \hyperlink{bib28}{2020}) and that this could potentially cause a rapid spread of the disease. 
It was then realized that public health measures such as masking and physical distancing would be necessary in ordinary situations since an infected person (symptomatic or asymptomatic) who merely engaged in a conversation could spread the disease (Asadi et al., \hyperlink{bib2}{2020}). More recently, infection through the airborne virus has been much in the news.
These findings highlight the need to better understand the different transmission modes of SARS-CoV-2, in which the fluid dynamics of droplet/aerosol transport plays a key role (Bourouiba, \hyperlink{bib8}{2021}). In particular, scientific attention has been focused on physical distancing guidelines as set out in public health advisories, which were based at first on a study done in the 1930s (Wells, \hyperlink{bib45}{1934}). According to that study, large droplets (>100$\mu$m) settle faster than they evaporate but can cover $\approx$1m horizontally while doing so, causing direct transmission.
It was estimated that smaller droplets cannot directly transmit the disease due to their fast evaporation. Recent research has shown that smaller droplets can cause disease as well. There is an entire spectrum of droplet sizes released during different expiratory events, and a significant part of this size spectrum can be transported over long distances by the turbulent jet/puff generated during such events (Bourouiba, \hyperlink{bib8}{2021}). The turbulent jet further contains water vapour which slows down evaporation.
 The complexity of the transmission dynamics of SARS-CoV-2, coupled with its asymptomatic transmissibility, could thus have played a large role in the ongoing pandemic afflicting the world (Prather et al., \hyperlink{bib31}{2020}), with several countries experiencing second or third waves of the contagion at present. 

Bourouiba (\hyperlink{bib8}{2021}) discusses the break-up of mucosalivary fluid bubbles and the formation of  droplets ranging from 1$\mu$m to 500$\mu$m (Duguid, \hyperlink{bib18}{1946}; Johnson et al., \hyperlink{bib24}{2011}) during different expiratory events. Among this range, the large virus-laden droplets predominantly cause direct transmission or contaminate surfaces close to their source (Basu et al., \hyperlink{bib5}{2020}; Bhardwaj \& Agrawal, \hyperlink{bib6}{2020}) turning them into “fomites”. Small droplets can stay airborne for longer times, their longevity being a function of their composition as well as prevalent ambient conditions like the relative humidity and temperature.
Furthermore, small droplets can evaporate completely while airborne and turn into what are called droplet nuclei. 
The SARS-CoV-2 has been found to survive in aerosols (typically consisting of micro-droplets and droplet nuclei of sizes less than 5$\mu$m) with a half-life of about 15 minutes for typical indoor conditions (Marr et al., \hyperlink{bib27}{2019}; US dept of Homeland Security, \hyperlink{bib30}{2020}; Schuit et al., \hyperlink{bib35}{2020}); another study (Van Doremalen et al., \hyperlink{bib43}{2020}) found the virus half-life in aerosols to be 1.1 hour (Greenhalgh et al., \hyperlink{bib21}{2021}). The long-range transport of disease-causing viruses through such droplets and droplet nuclei, and the impact of the ambient conditions on virus survival, are open questions of fluid dynamical interest (Bourouiba, \hyperlink{bib7}{2020}), with some studies suggesting that droplets with diameters in the range of (2.5$\mu$m-19$\mu$m) have the greatest potential for causing the initial nasopharyngeal infection (Basu, \hyperlink{bib4}{2021}; Smith et al., \hyperlink{bib38}{2020}). Droplets exhaled while speaking are typically smaller than 10$\mu$m (Asadi et al., \hyperlink{bib3}{2019}) and lie in the described range for higher nasopharyngeal infection.

Most of the fluid dynamical studies on human expiratory flows have focused on droplet transport during coughing and sneezing, due to their direct relevance to symptomatic transmission. Several
experimental (Bourouiba et al., \hyperlink{bib9}{2014}; Clark \& De Calcina-Goff, \hyperlink{bib14}{2009}; 
Gupta et al., \hyperlink{bib22}{2009}; Nielsen et al., \hyperlink{bib29}{2012}; Tang et al.,\hyperlink{bib41}{2009}; Wei \& Li, \hyperlink{bib44}{2017})
and numerical (Dbouk \& Drikakis, \hyperlink{bib16}{2020}; 
Fabregat et al., \hyperlink{bib19}{2021}; Liu et al.,\hyperlink{bib26}{2017}; Qian {\&} Li, \hyperlink{bib32}{2010}) 
studies of cough and sneeze flows have been reported and their outcomes are being incorporated into epidemiological models (Chaudhuri et al., \hyperlink{bib10}{2020}; Dbouk \& Drikakis, \hyperlink{bib17}{2021}).
On the other hand, the fluid dynamics of speech and breathing has received much less attention until very recently. Shao et al., \hyperlink{bib36}{2021}
provide a risk assessment of virions exhaled due to normal breathing in elevators, classrooms, and supermarket settings. Chong et al., \hyperlink{bib12}{2021} carried out a direct numerical simulation (DNS) of a turbulent vapour puff
and found that droplets can last O(100) times longer in cold humid air than predicted by classical models (Wells, \hyperlink{bib45}{1934});
they used an inlet flow-rate profile typical of a cough but the results are equally applicable to other expiratory flows including speech. Recent experiments (Giovanni et al., \hyperlink{bib20}{2021}) on the effects of airflow velocity on droplet trajectory in speech and vocal exercises have been used to construct models of the behavior of different droplets. Abkarian et al.(\hyperlink{bib1}{2020})
carried out a large eddy simulation of speech flows generated by the repetition of certain phrases. They showed that beyond a certain distance from the mouth of the speaker, speech flow behaves like a steady jet that spreads at a typical half-angle of $10^\circ$.
These results were incorporated by Yang et al. (\hyperlink{bib49}{2020}) into a simplified model for the transport of aerosol particles away from a speaker's mouth, using known properties of steady jets like the $1/x$ variation of the velocity and scalar concentration with the axial distance $x$. They calculated the probability of infection of a silent listener based on conservative estimates of the minimum number of virions $(N_{inf}\approx 100)$
that must be inhaled to cause infection (Basu, \hyperlink{bib4}{2021}; Kolinski \& Schneider, \hyperlink{bib25}{2020}) and presented space-time maps of the risk of infection. This study highlighted the fact that the disease transmission by speech involves not just distances but also exposure times (see also Tan et al., \hyperlink{bib40}{2021}) and that these should also be incorporated in the public health guidelines.

The idealized scenario studied so far, where a single person engages in an extended monologue, is of limited applicability, and typical public interactions are dialogues of short time-spans, e.g., over- the-counter conversations at a supermarket. The present study is devoted to gauging infection probability from such a short, unmasked conversation between two people, by performing a DNS of speech flow. We compute turbulent transport of speech aerosols, which play a key role in the transmission of virus. We estimate the total viral ingestion by a listener by including exposure through the eyes and mouth, in addition to the aerosols inhaled through the nose (more commonly only the last is used to determine viral dose). We show that the active participation in the conversation of the second person significantly alters the evolution of the jet from the first person, dramatically mitigating infection probability. Any temporal asymmetry in speech enhances the risk of infection to the person who speaks less. Secondly and rather interestingly, a small vertical or lateral separation between the mouths of the speakers actually increases the infection probability, due to a less effective interference of the two jets. At large vertical or lateral separation, infection probability is lower, as would be expected. We discuss the implications of these results for improving physical distancing guidelines and for epidemiological modelling. Also, our results offer interesting experimental test cases and can be used to validate flow-modelling approaches such as the Reynolds-averaged Navier Stokes equations.

The paper is organized as follows. First, we describe the computational geometry and set-up (Section 2), followed by the parameters needed to estimate the infection risk. We present two cases: one in which one of the people is a passive listener, and the other in which both people converse. Next, we describe our method of determining the total exposure and ingestion of virions, which is used to calculate the probability of infection for the above two cases (Section 3).  Lastly, a case of conversation with temporal asymmetry in speech duration is analyzed before a discussion (Section 4) and summary (Section 5) of the results.

\section{Numerical Experiments}

\begin{figure}[!h]
\centering%
\begin{minipage}{0.38\textwidth}
  \includegraphics[width=1.0\textwidth]{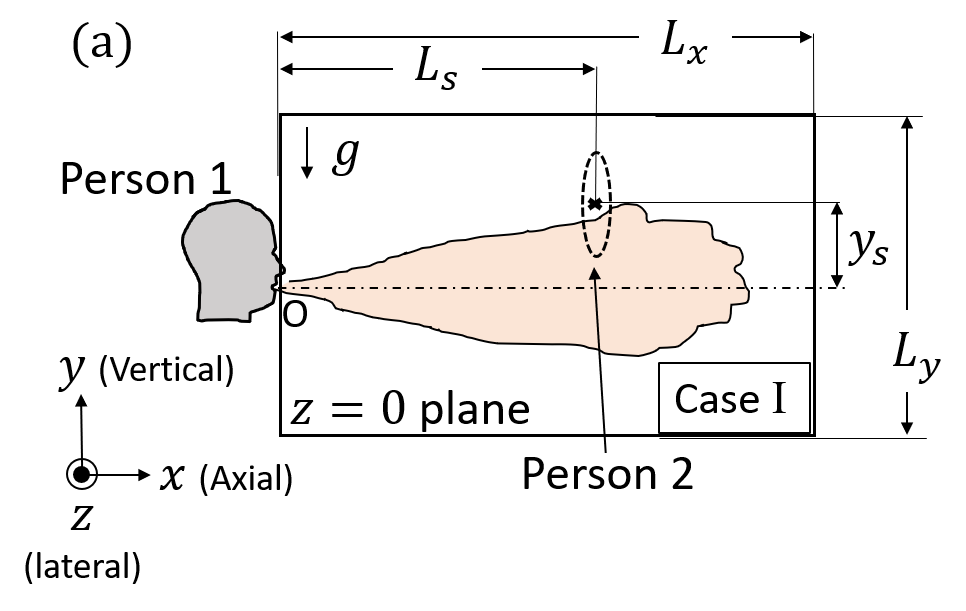}
\end{minipage}%
\hspace{0.05\textwidth}%
\begin{minipage}{0.52\textwidth}
  \includegraphics[width=1.0\textwidth]{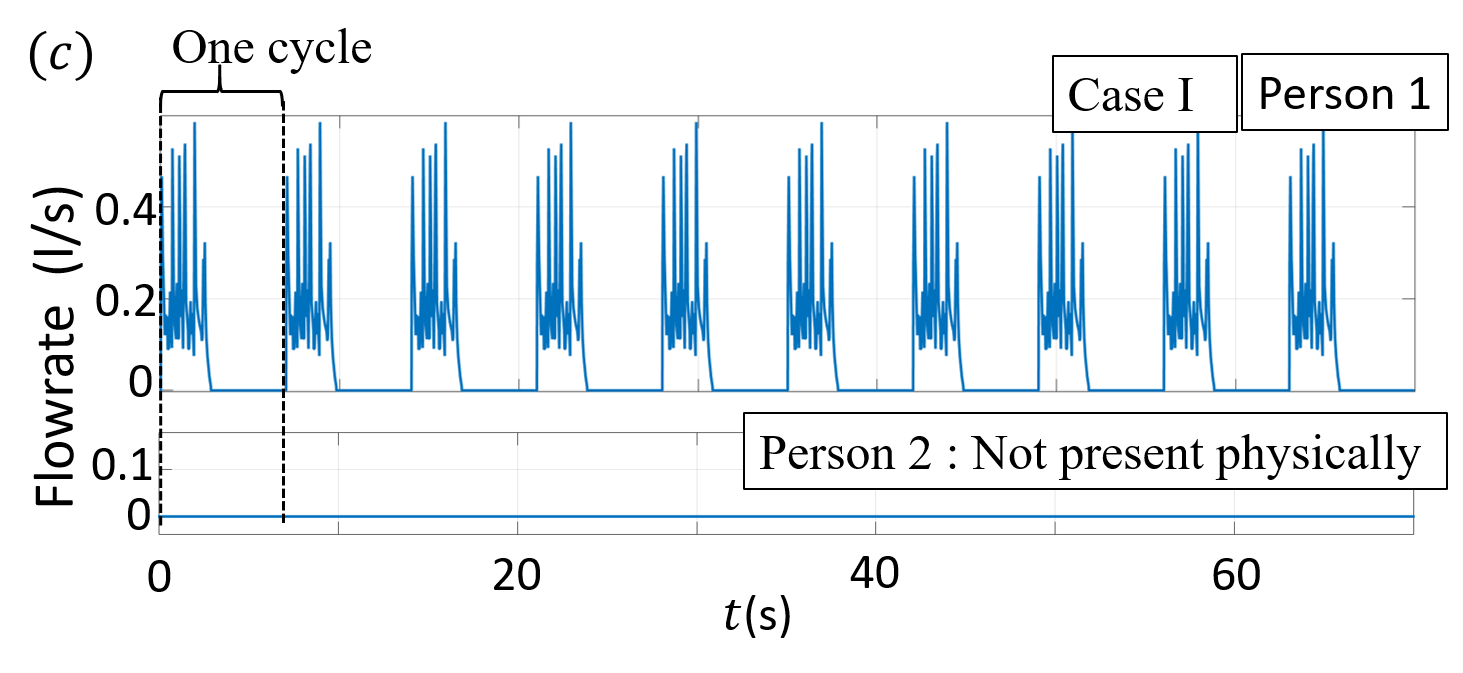}
\end{minipage}\\
\begin{minipage}{0.39\textwidth}\centering
  \includegraphics[width=1.0\textwidth]{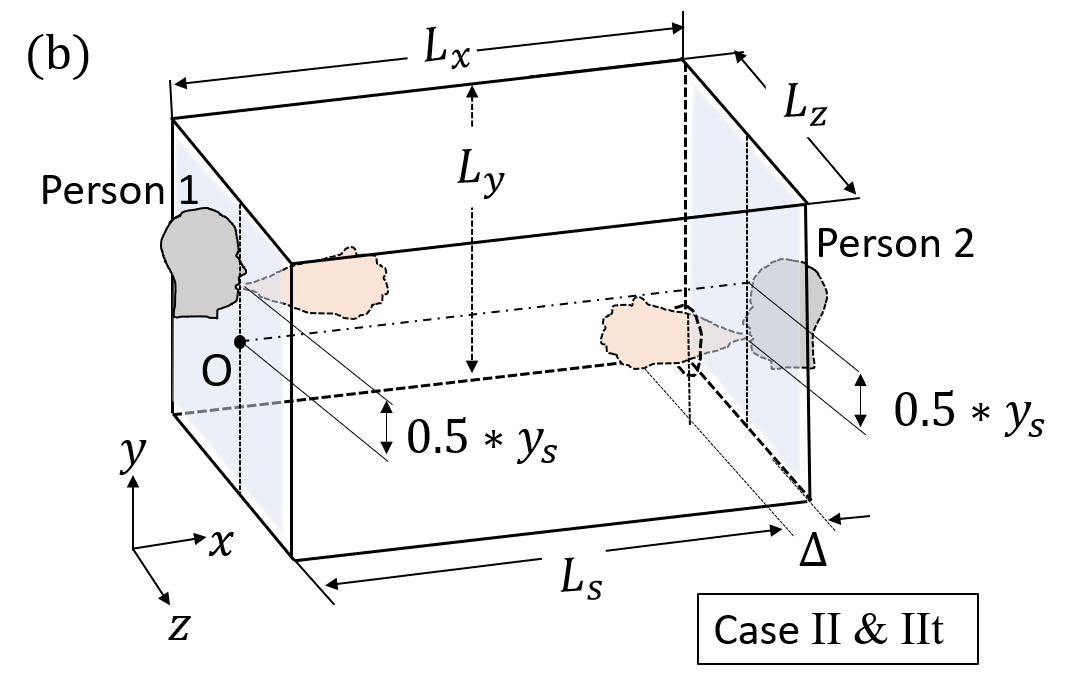}
\end{minipage}
\hspace{0.05\textwidth}%
\begin{minipage}{0.52\textwidth}\centering
  \includegraphics[width=1.0\textwidth]{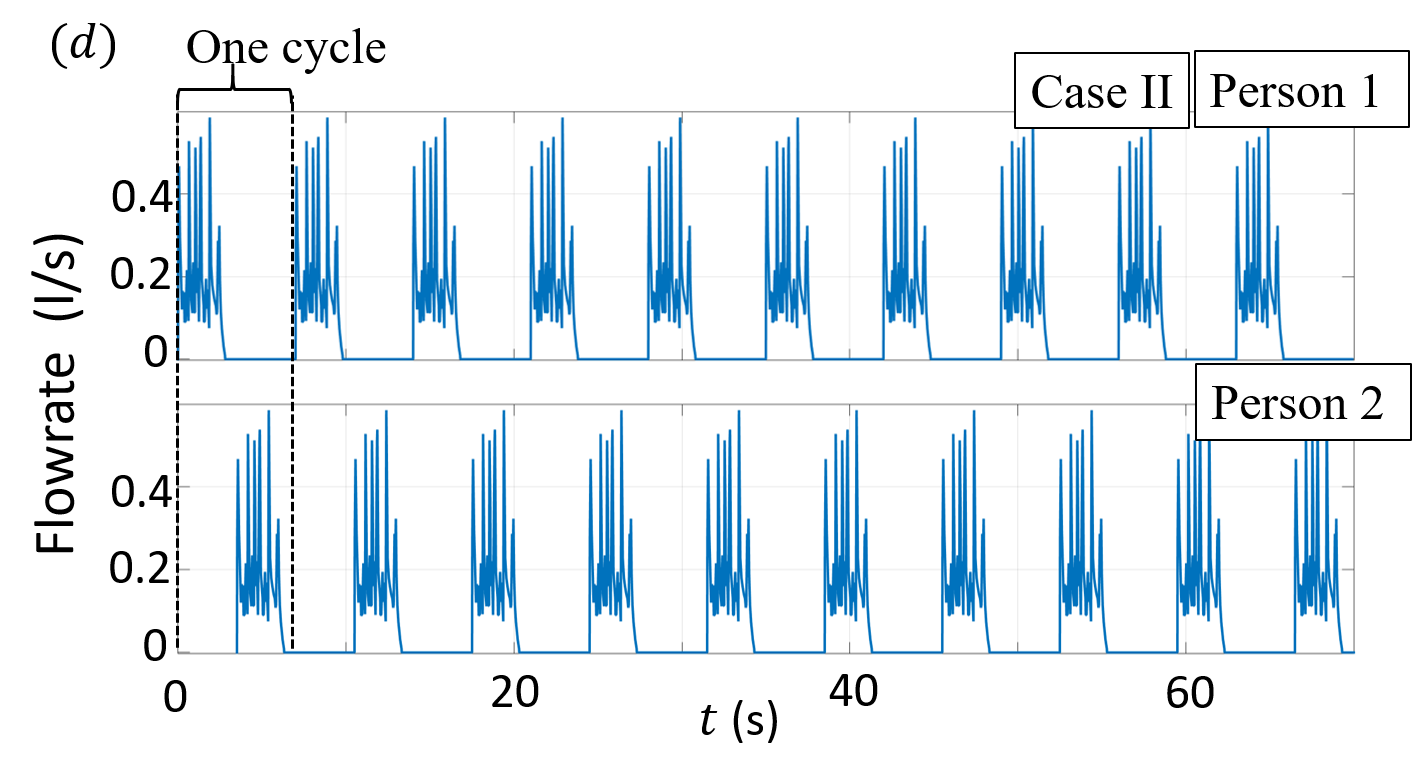}
\end{minipage}
\caption{(a) Side view $(z=0)$ of the computational domain in Case I showing the location $(L_s,y_s)$ of the silent (and susceptible) listener, represented by a circular region of interest (ROI) as shown in the figure. (b) Three-dimensional representation of the computational domain in Case II, where both Person 1 and Person 2 speak, and $L_s$ and $y_s$ are treated as parameters. The orifices are displaced symmetrically about $y=0$. (c) Inlet flow rate (in $l/s$) from the orifice `O' in Case I. (d) Inlet flow rates from the two orifices in Case II.\label{fig1}}
\end{figure}

We use two types of computational domain for the direct numerical simulation of speech flows. The first type of simulation, Case I, is done using a domain whose side view is shown in \hyperref[fig1]{Figure~\ref{fig1}a} (the lateral size of the domain is the same as that shown in \hyperref[fig1]{Figure~\ref{fig1}b});
the co-ordinate system used is shown in the figure.
Here, a silent susceptible listener is prescribed at $(L_{s},y_{s},z=0)$ 
with respect to the origin 'O' where the speaker (Person 1) is placed. A circular region of interest (ROI) of diameter 17.2cm (which is approximately the diameter of a human face) is used to represent the face of Person 2 (\hyperref[fig1]{Figure~\ref{fig1}a}). To examine the 6ft rule, the length of the computational domain is chosen to be $L_{x}=1.96$m,
and the vertical and lateral extents are chosen to be $L_{y} =L_{z} =0.98$m
which are sufficient for the study due to the small spread angle $(\approx20^\circ -28^\circ)$ of speech flows (Yang et al., \hyperlink{bib49}{2020}).
The grid resolution used is $256\times128\times128$ in $L_x\times L_y\times L_z$, 
which is equivalent to $256^3$ for a cubical domain with size $L_x$; 
this resolution has been tested for grid convergence by using higher resolutions of $512^3$ and $1024^3$(see Fig. S1).
As done in previous studies (Ravichandran \& Narasimha, \hyperlink{bib40}{2020}; Singhal et al., \hyperlink{bib37}{2021}), open boundary conditions are imposed on the four lateral boundaries at $y=0,L_y$ and $z=0,L_z$
and the boundary at $x=L_x$;
this enables advection of the turbulent structures out of the domain, without affecting the upstream flow. 
A free-slip condition is used for the boundary at $x=0$, except for the orifice at `O' (see below).

Case II is a set of simulations where the two people are in conversation (\hyperref[fig1]{Figure~\ref{fig1}b}), and the flows thus generated interact with each other.
We account for a possible height difference between the speakers by a vertical separation $y_s$. 
The orifice for Person 1 is located at $(0,0.5*y_s,0)$ and that for Person 2 (whose susceptibility to infection is being estimated) at $(L_s+ \Delta ,-0.5*y_s,0)$, as shown schematically in \hyperref[fig1]{Figure~\ref{fig1}b}. 
To account for the obstacle created by Person 2 to the incoming speech flow for this case, the boundary at $x=L_x=(L_s  +\Delta)$ , excluding the orifice, is prescribed to be a free-slip wall (instead of an outflow boundary) similar to that at $x=0$.
In order to avoid wall effects, we choose an axial location that is at a distance $\Delta (\approx 0.15m)$ upstream of $L_x$ to place the ROI.
The axial length of domain is varied among three values, $L_s\in\{0.6\rm{m},1.2\rm{m},1.8\rm{m}\}$ , keeping the lateral lengths $L_y=L_z=0.98$m fixed.
We also wish to understand the rate of infection when the conversation is temporally asymmetric, i.e., Person 1 speaks more than Person 2. 
We simulate such situations with $L_s=1.2$m and $y_s=0$ and name it “Case IIt”.

The speaker’s mouth is modelled as an elliptical orifice with half-axis dimensions of $a=1.5\rm{cm},b=1.0\rm{cm}$ , following Abkarian et al.(\hyperlink{bib1}{2020}).
We use the characteristic diameter $d=2\sqrt{ab}=2.45cm$ of the ellipse as the length-scale for non-dimensionalisation; 
the non-dimensional size of the domain is therefore $(80d\times 40d\times 40d)$ for $L_s=1.8$m.
Prescribing speech flow at the orifice is not straightforward as it involves complex linguistic expression, which can continue for an extended period of time. 
Here we follow Abkarian et al.(\hyperlink{bib1}{2020}) in choosing a simple repetitive phrase for simulating speech flow.
Abkarian et al.  (\hyperlink{bib1}{2020}) show that speech phrases with plosive sounds (e.g., words containing the letter “P” - oshtavya in Sanskrit) induce puffs that travel further than those having fricative sounds (e.g., the letter “S”).
Furthermore the speech jet produced by plosives was found to be directed primarily in the axial direction  (Abkarian et al., \hyperlink{bib1}{2020}), obviating the need to introduce flow directionality at the orifice.
We therefore use a repeated utterance of the phrase "Peter Piper picked a peck" and prescribe the inlet velocity at the orifice (in the axial direction) 
using laboratory measurements of flow velocity associated with this phrase (see Figure 4c of  Abkarian et al., \hyperlink{bib1}{2020}).
We do not include the inhalation part of the cycle used in  Abkarian et al. (\hyperlink{bib1}{2020}) as it does not affect the transport of aerosols, except for a region very close to the orifice  (Abkarian et al., \hyperlink{bib1}{2020}).
In the present study, the exhalation phase of the phrase has a volume of 0.5 litre per cycle, as in Abkarian et al.  (\hyperlink{bib1}{2020}), which lasts for 2.8s, followed by a halt in speaking of 4.2s, which makes one complete cycle of speech (\hyperref[fig1]{Figure~\ref{fig1}c and d}).
The present simulations are validated against the results of Abkarian et al.  (\hyperlink{bib1}{2020}), and show a good overall agreement.
The cycle-averaged centreline velocity from our simulations reproduces the $1/x$ variation after $x\approx0.4$m (see Fig. S1) and the “flow length” exhibits the  $t^{1/4}$ variation for small $t$ (typical of a “puff”), followed by the $t^{1/2}$ variation at large $t$ (typical of a “starting jet”) as obtained in Abkarian et al. (\hyperlink{bib1}{2020}) (Fig. S2).

For Case I, we have used 10 speech cycles for Person 1 and zero flow rate for Person 2 (who acts like a silent listener, though not physically present in the domain); see \hyperref[fig1]{Figure~\ref{fig1}a and c}.
For the simulations in Case II, flow rates at both orifices (corresponding to Persons 1 and 2) consist of 10 staggered cycles shown in \hyperref[fig1]{Figure~\ref{fig1}d};
this represents each person taking turns to speak the same phrase.
In Case IIt, Person 1 and Person 2 have 10 cycles of conversation, following which, Person 2 stops speaking, and Person 1 continues for another 10 cycles (see \hyperref[fig7]{Figure~\ref{fig7}}).
For this case, the total speech time (including the silent intervals) is 140s for Person 1 and 70s for Person 2. 
The maximum flow velocity at the orifice, $u_c=1.167m/s$, is used as the velocity scale for all the cases.
Since these flow speeds are small compared to the speed of sound in air and also because the change in density within the flow is much smaller than the ambient density $(|\Delta\rho|⁄\rho\approx0.05$ for $\Delta T=14^\circ C$ at the orifice; see appendix~\ref{appendix_A})
we may make the Boussinesq approximation.
This involves treating the velocity field as effectively incompressible with the density differences relevant only for the buoyancy term in the momentum equation.
The buoyancy module of our code has been validated for a steady plume (Singhal et al., \hyperlink{bib37}{2021}) wherein the self-similar decay of the centreline temperature has been accurately captured.
For the speech flows considered here, it turns out that the buoyancy forces are much smaller than the inertia forces.
As a result, the flow does not show any perceptible deflection in the vertical direction on average (Fig. S3; see also \hyperref[fig4]{Figure~\ref{fig4}b}) and is qualitatively similar to that reported in  Abkarian et al. (\hyperlink{bib1}{2020}) who did not solve for temperature . 
This is borne out by the small value of $Fr^{-2} (=0.00842)$ obtained for the present simulations based on the orifice conditions, where $Fr$ is the Froude number ({appendix~\ref{appendix_A}}).
Furthermore, the temperature drops rapidly as the speech jet issues from the orifice (by about $10^\circ$ C over 0.5m from the orifice; see Fig. S4) and once the steady “self-similar” regime is established for $x>0.45$m Fig.( S5b), the Froude number based on local parameters is expected to remain constant (Turner,\hyperlink{bib42}{1986}). 
Thus, buoyancy effects are estimated to be negligible over the entire flow evolution.

In order to model the dynamics of virus laden droplets generated during speech, we make two simplifications.
First, that most droplets ($O(10\mu \rm{m})$ or less) produced during speech are small enough that they follow fluid streamlines.
The smallness can be quantified in terms of the Stokes number of the droplets,
defined as the ratio $\tau_p/\tau_f$, where $\tau_p$  is the timescale on which the velocity of a droplet adjusts itself to the local fluid velocity,
and $\tau_f=d/u_c$ is the flow timescale.
For droplet sizes typical of speech flows, Stokes numbers are much smaller than 1 (Yang et al., \hyperlink{bib49}{2020}),
which means that within a fraction of the flow timescale, a droplet attains the same velocity as the flow. In other words, it behaves as a passive scalar. In the present work, $\tau_p\sim 0.5$ms (for $10\mu \rm{m}$ droplets) and $\tau_f \sim 20$ms, giving $\tau_p/\tau_f\sim 0.025$ which is much less than 1. Since the flow velocity rapidly drops away from the orifice, the passive-scalar approximation is  justified in the entire domain. This approach has been used before for studying the interaction of droplets and turbulence in a cloud flow (Ravichandran \& Narasimha, \hyperlink{bib34}{2020})
and allows us here to represent small droplets in speech flow by scalar fields of concentration $C_{s1}$ and $C_{s2}$ emanating from Persons 1 and 2 respectively (for Case I, $C_{s2}=0$).
We prescribe unit concentration at the orifices during speaking and 0 otherwise. Secondly, we note (a) that the total liquid content in these flows is quite small (Asadi et al.,\hyperlink{bib3}{2019}), as are the temperature changes resulting from droplet evaporation;
and (b) that the half-life of SARS-CoV-2 in speech droplets/nuclei ($\approx$ 15 mins to 1 hr) is much longer than the time-scales of evaporation (a few seconds; Chaudhuri et al., \hyperlink{bib10}{2020}; Yang et al., \hyperlink{bib49}{2020}).
Thus, even if the liquid part of a droplet has evaporated, the nucleus containing the infection survives. We therefore ignore the thermodynamics of evaporation in this study, and thus treat small droplets and droplet nuclei as interchangeable, consistent with Abkarian et al. (\hyperlink{bib1}{2020}) and Yang et al. (\hyperlink{bib49}{2020}).
We refer to both as “aerosols”.
The resulting governing equations, namely the Boussinesq Navier-Stokes equations, the continuity equation and the scalar transport equations, are solved using the finite difference DNS solver, Megha-5 ({for further details, see appendix~\ref{appendix_A}}).

\section{Results}
\subsection{Case I: Person 1 speaking and Person 2 being a silent listener}

\begin{figure}[!h]
\centering%
\begin{minipage}{0.45\textwidth}
  \includegraphics[width=1.0\textwidth]{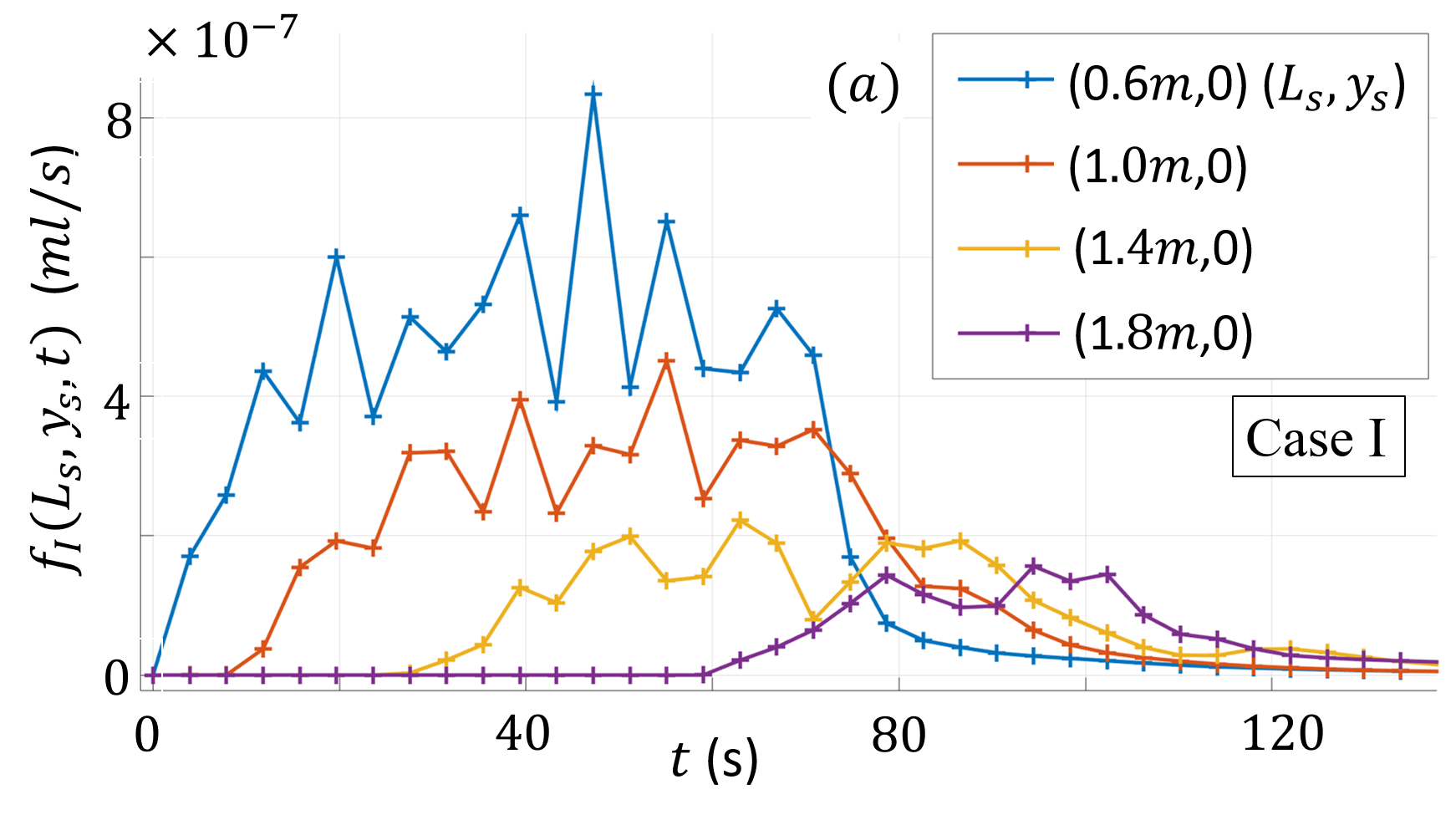}
\end{minipage}%
\hspace{0.1\textwidth}%
\begin{minipage}{0.45\textwidth}\centering
  \includegraphics[width=1.0\textwidth]{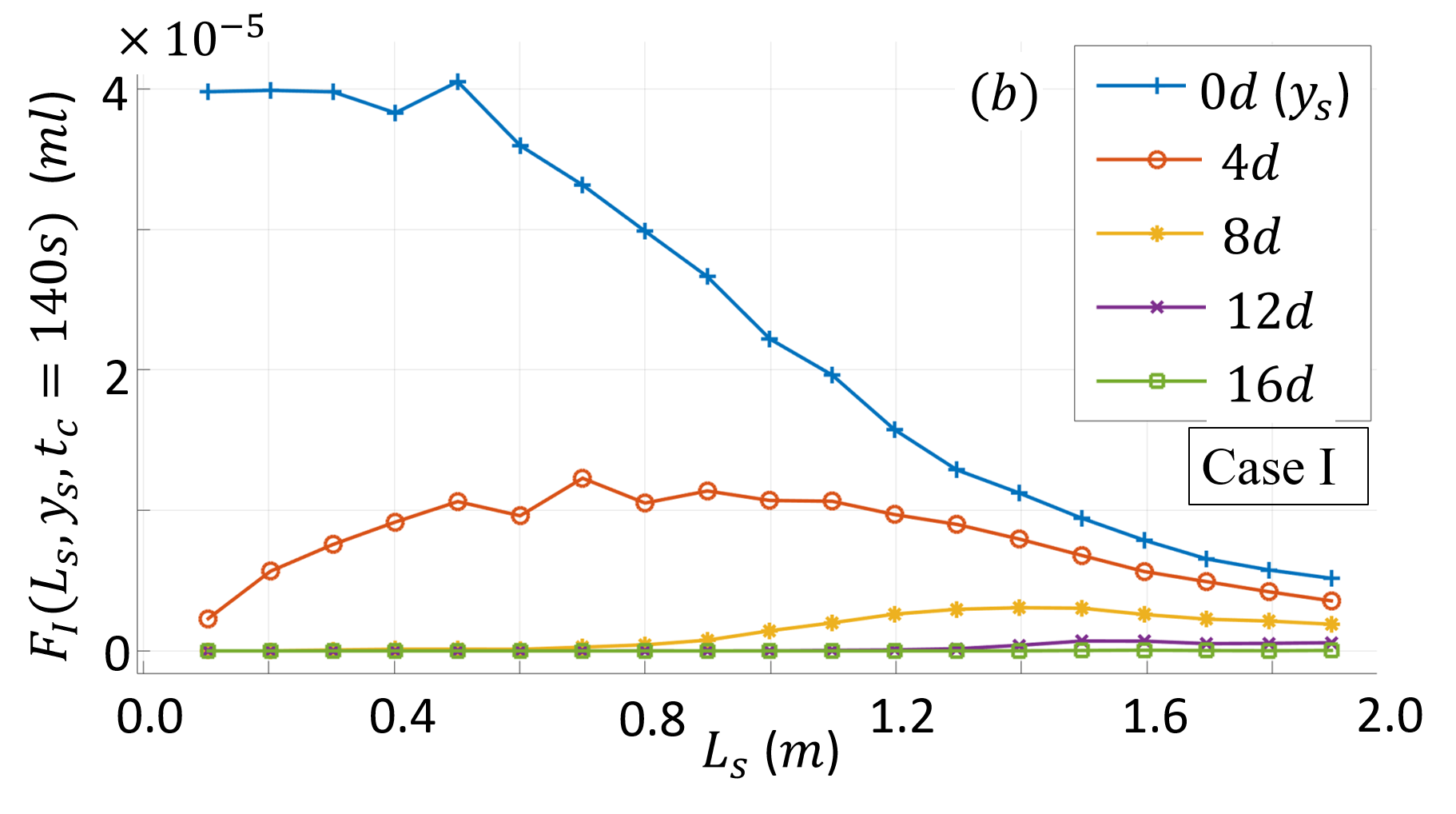}
\end{minipage}\\
\begin{minipage}{0.44\textwidth}
  \includegraphics[width=1.0\textwidth]{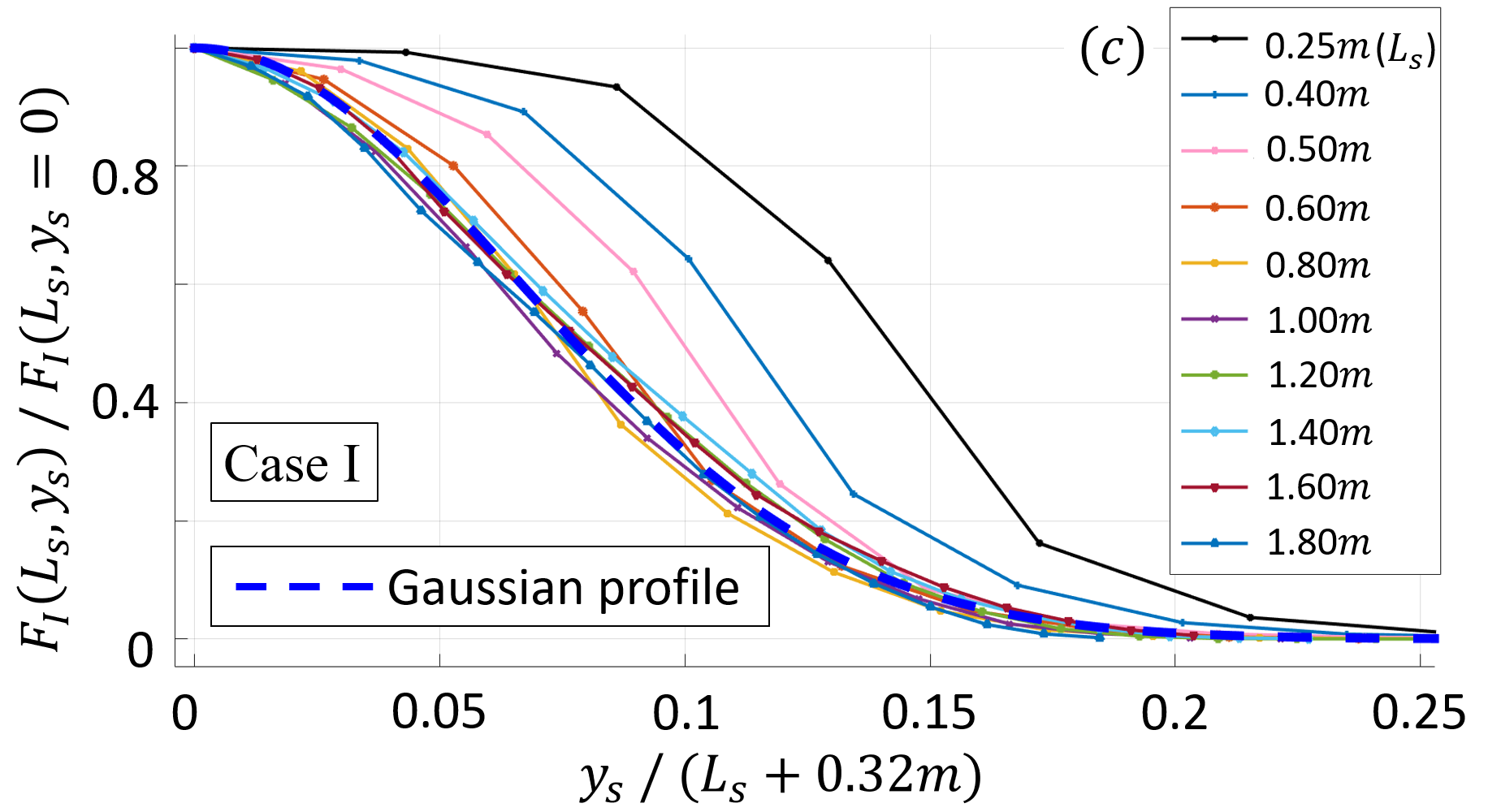}
\end{minipage}%
\caption{Aerosol flux for Case I.  (a) Variation in time of the aerosol flux $f_I$ through the circular ROI centred at $(L_s,y_s=0)$. Close to the orifice, the flow is resembles a series of puffs rather than a jet. (b) The total scalar exposure $F_I (L_s)$ over the simulation time (140s), for different height differences $y_s$. Interestingly, $F_I$ varies non-monotonically with $L_s$ for non-zero $y_s$. This is because only a part of the speech-flow jet intersects the ROI for $L_s   \lesssim 10y_s$.  (c) Vertical variation of normalized $F_I (L_s,y_s )$ and its comparison with a Gaussian curve. The normalized $F_I$ profiles show best collapse with a virtual origin at 0.32m upstream of the orifice.\label{fig2}}
\end{figure}

Studies (Chaudhuri et al., \hyperlink{bib10}{2020}; Yang et al., \hyperlink{bib49}{2020}) have provided spatial and temporal separation guidelines using calculations based on the total viral load on a susceptible person present at a specific location. 
The infection probability is then calculated from the number of virions inhaled by the person.
However, several studies have found that infections may also be caused by viral contact through other exposed areas like the eyes (Chen et al., \hyperlink{bib11}{2020}; Coroneo \& Collignon, \hyperlink{bib15}{2021}; Sun et al., \hyperlink{bib39}{2020}; Xie et al., \hyperlink{bib48}{2020}) or the mouth and lips (WHO scientific brief 27 March, \hyperlink{bib46}{2020}).
In order to include these possibilities, we estimate the exposure of a listener to virus-laden aerosols over an ROI positioned in front of their face.
Towards this, we determine aerosol concentration $(\phi)$ which is related to the passive scalar concentration as $\phi=\phi_o$  ($C_s/C_{so}$), where  $\phi_o$ represents the volume of droplets (i.e., liquid volume)
per unit volume of air at the orifice and $C_{so}=1$ ({see appendix~\ref{appendix_A}}).
We specify $\phi_o=6\times10^{-9}$ (Yang et al., \hyperlink{bib49}{2020})  as a representative value. As mentioned earlier, the speech droplets evaporate fast and turn into
droplet nuclei but the total number of virions carried by the speech aerosols remains the same; see Yang et al. (\hyperlink{bib49}{2020}). The flux of aerosols through an ROI centered at $(L_s,y_s,z=0)$ is given as 

\begin{equation}
    f(L_s,y_s,t)=\int\int\phi_1\ u\ dy\ dz,\quad  \quad 2\sqrt{{(y-y_s) }^2+z^2} <D_{ROI}\label{eq1}\
\end{equation}
where $\phi_1$ is the aerosol concentration corresponding to the speech flow of Person 1, $u$ is the axial velocity and $D_{ROI}=17.2$cm is the diameter of the ROI. 
The aerosol flux through the ROI is taken as the viral exposure to the face of the listener.
Since the deflection of the flow by buoyancy is negligible, the $y$ and $z$ directions are equivalent. We therefore only need to vary $L_s$ and $y_s$.
\hyperref[fig2]{Figure~\ref{fig2}a} shows the time series of the aerosol flux across an ROI for Case I, $f_I$ $(L_s,y_s=0,\ t)$, using the data obtained from the simulations.
The transition of the flow from a puff-like behavior to jet-like behavior can be seen from the decrease in the oscillations of the scalar flux with increase in $L_s$. The total aerosol exposure to the listener $F(L_s,y_s,t_c )$ over time $t_c$ is given by the time integral of the aerosol flux through the ROI,

\begin{equation}
    F(L_s,y_s,t_c) = \int_0^{t_c}f(L_s,y_s,t)dt,  \label{eq2}\
\end{equation}
where $t_c$ is the duration of exposure and $F$ is units of volume (here ml). Here we show results for  $t_c=140$s (although the total speech duration is 70s)
as this represents the time until which the listener at $L_s=1.8$m continues to receive aerosol flux from the speaker (\hyperref[fig2]{Figure~\ref{fig2}a and b}). 
The aerosol exposure $F_I (L_s,y_s,140s)$ for Case I is plotted in \hyperref[fig2]{Figure~\ref{fig2}b} as a function of axial location $L_s$ at five different vertical locations $y_s$.
For $y_s=0$, the total exposure is practically constant for $L_s<0.5$m, because the area of the ROI is larger than the cross-section of the jet till this axial location. Once the jet cross-section area exceeds the area of the ROI, the total exposure starts decreasing (\hyperref[fig2]{Figure~\ref{fig2}b}).
When the listener is not aligned with the speaker face to face $(y_s\neq 0)$, $F_I$  increases with $L_s$ for $L_s   \lesssim 10y_s$, before decaying at large $L_s$. 

The vertical profiles of $F_I (L_s,y_s,140\rm{s})$ are found to be bell-shaped curves with the peak in $F_I$  decreasing with increase in $L_s$. Suitably scaled versions of the profiles of $F_I$ are plotted in \hyperref[fig2]{Figure~\ref{fig2}c}.
The $F_I$ profiles at  $L_s=0.25\rm{m}, 0.40\rm{m}$, and $0.50\rm{m}$, show an evolution in shape, and show an approach towards a Gaussian distribution (represented by a dashed curve); this is consistent with the observation that flow is in a transition state between a puff and steady jet for these distances.
For $L_s> 0.50$m, the curves collapse well onto the Gaussian curve (\hyperref[fig2]{Figure~\ref{fig2}c}). Note that the Gaussian distribution for the scalar flux profiles is typical of a self-similar jet (Turner, \hyperlink{bib42}{1986}).
A more commonly used indicator for the start of the self-similar regime is the $1/x$ variation of the centreline scalar concentration. Such a variation starts from $L_s \approx 0.45$m (Fig. S1 and S5b) and we find it convenient to associate this location with the beginning of steady, self-similar speech flow.
These results are useful for determining viral exposure to the listener’s eyes and mouth, as will be seen.

\begin{figure}[!h]
\centering%
\begin{minipage}{0.49\textwidth}
  \includegraphics[width=1.0\textwidth]{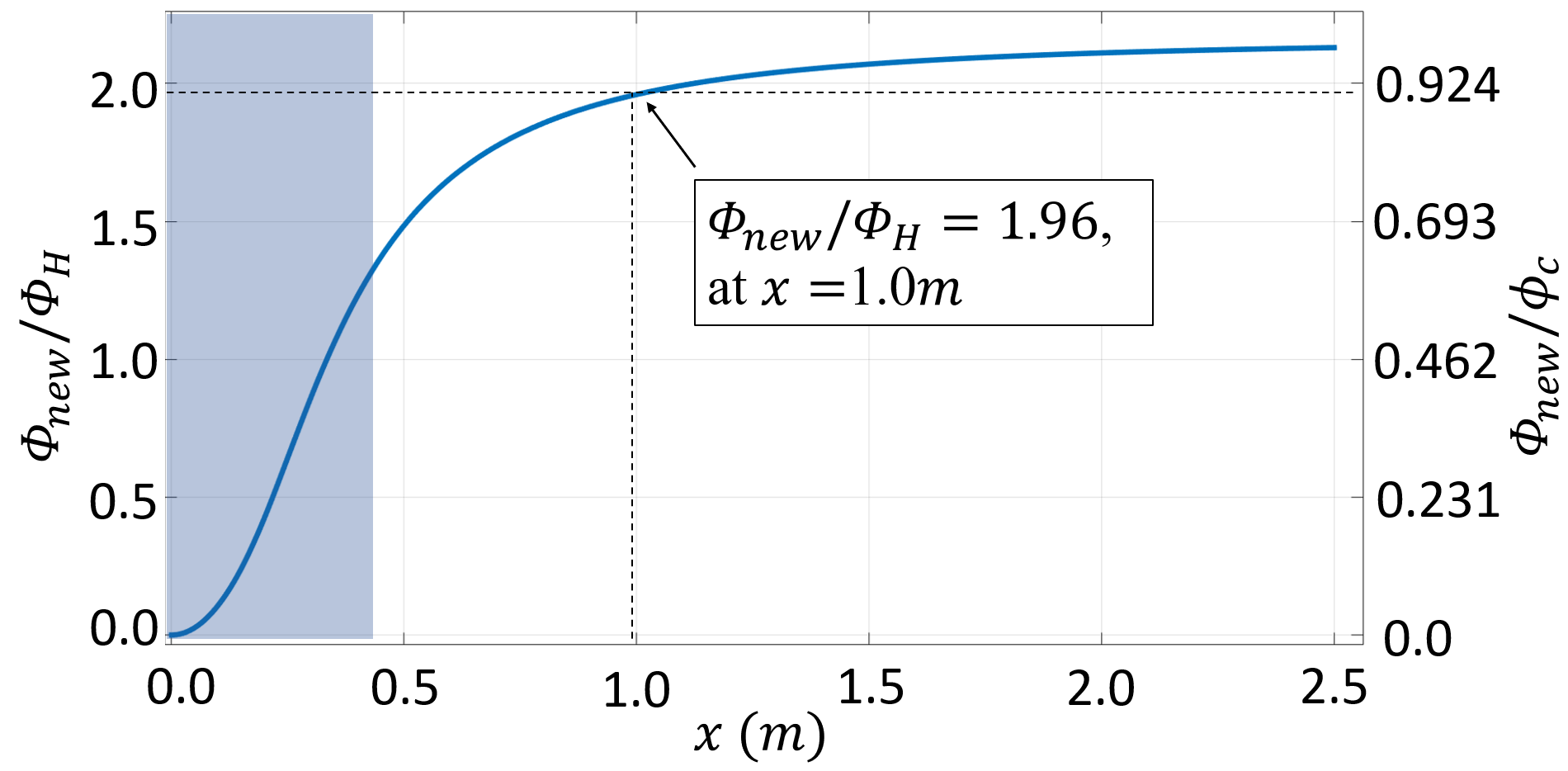}
\end{minipage}%
\caption{Comparison of the different characterizations of aerosol concentration. The axial variation of locally averaged aerosol concentration $(\Phi_{new})$ calculated by equation (\ref{eq4})  and its comparison with aerosol concentration based on the “top-hat” formulation $(\Phi_H)$. Steady self-similar behavior is established only for $L_s \gtrapprox 0.45m$. \label{fig3}}
\end{figure}

We first obtain a conservative estimate of the aerosol flux inhaled by a listener by using the fact that the domain of inhalation is much more localized than has been considered in some of the previous studies. 
In particular, Yang et al. (\hyperlink{bib49}{2020}) have used a measure of aerosol concentration ($\Phi_{H}$) that is averaged over an area with radius $r_H=x tan\alpha$, where the half-cone angle $\alpha$ encompasses 90\% of scalar within it.
These effectively represent “top-hat” quantities and therefore have been denoted by a subscript “H”. Using this formulation, Yang et al.  estimated the number of inhaled virions as
 \begin{equation}
     N(\Phi_H )\ =\ c_v \Phi_H (x)Q_r t=c_v \phi_o a_H Q_r t/(x tan\alpha),\label{eq3}\ 
 \end{equation}
where $c_v$ is the number of virions per unit droplet volume (or the total saliva volume) in the exhaled air during speech, $a_H$ is the orifice radius and $Q_r$ is the rate of inhalation by the listener taken as $0.1 l/s$.
However, the actual distributions of axial velocity and scalar concentration in a steady self-similar jet are Gaussian, and it is therefore of interest to relate $r_H$ and $\Phi_H$ with their Gaussian counterparts.
The relevant length scale characterizing a Gaussian distribution is the “$1/e$ width” $(b_e)$, defined as a radial distance where a quantity reaches $1/e$ times its centreline  value - for axial velocity we denote it as $b_{ue} (x)$ and for aerosol concentration as $b_{\phi e }(x)$.
The centreline values of the mean axial velocity and aerosol concentration are denoted as $U_c (x)$ and $\phi_c (x)$ respectively. We get the relation between the top-hat width of Yang et al. (\hyperlink{bib49}{2020})  and the Gaussian widths as $r_H=1.516b_{\phi e}$ $(x)=1.819b_{ue} (x)$  ({see\textit{ SI Text, section 1)}}.
Thus, the top-hat radius considerably overestimates the lateral spread of the aerosol distribution. 
As a result of this, the top-hat velocity $(v_H (x))$ and aerosol concentration $(\Phi_H (x))$ are underestimated in relation to their Gaussian counterparts as $v_H$ $(x)=0.39U_c (x)$ and $\Phi_H (x)=0.462\phi_c (x)$; (see Fig. S5 for a graphical comparison).
Thus, a susceptible listener can be expected to get exposed to a lot more number of virions than estimated in Yang et al.(\hyperlink{bib49}{2020}).
To make a realistic estimate of the aerosol flux inhaled by a listener, we consider inhalation to be a “sink” flow (with nose at its centre), drawing in air from a hemispherical domain with a radius of $6.2$cm (Abkarian et al., \hyperlink{bib1}{2020}; Haselton \& Sperandio, \hyperlink{bib23}{1988}). An average aerosol concentration is then calculated over this localized circular region centered at the jet axis (rather than over the much larger area, $\pi r_H^2$) given as

\begin{equation}
    \Phi_{new}=(b_{\phi e}/6.2)^2    [1-exp(-(6.2/b_{\phi_e})^2)]    \phi_c,\label{eq4}
\end{equation}
which represents the average scalar concentration a listener is likely to ingest through the inhalation process; here $b_{\phi e}$ is measured in cm ({for derivation, see SI Text, section 2}).
Using $\Phi_H (x)=0.462 \phi_c (x)$ and $b_{\phi e}=0.137x$ for a steady self-similar jet (Singhal et al., \hyperlink{bib37}{2020}), we obtain a relation between $\Phi_{new}$ and $\Phi_H$ as a function of $x$.
This is plotted in \hyperref[fig3]{Figure~\ref{fig3}}, which shows that $\Phi_{new}$  > $\Phi_H$   for $x > 0.45m$;
the region up to this axial location is shown shaded as the Gaussian profile does not apply in this region. The figure also shows that for $x>1\rm{m}, \Phi_H (x) \approx 0.5 \Phi_{new} (x)$; the use of the top-hat profile thus underestimates the risk of infection by direct inhalation of virions by about $50\%$, supporting our expectation mentioned earlier.

\subsection{Case II: Persons 1 and 2 engaged in conversation}

\begin{figure}[!h]
\centering%
\begin{minipage}{0.46\textwidth}
  \includegraphics[width=1.0\textwidth]{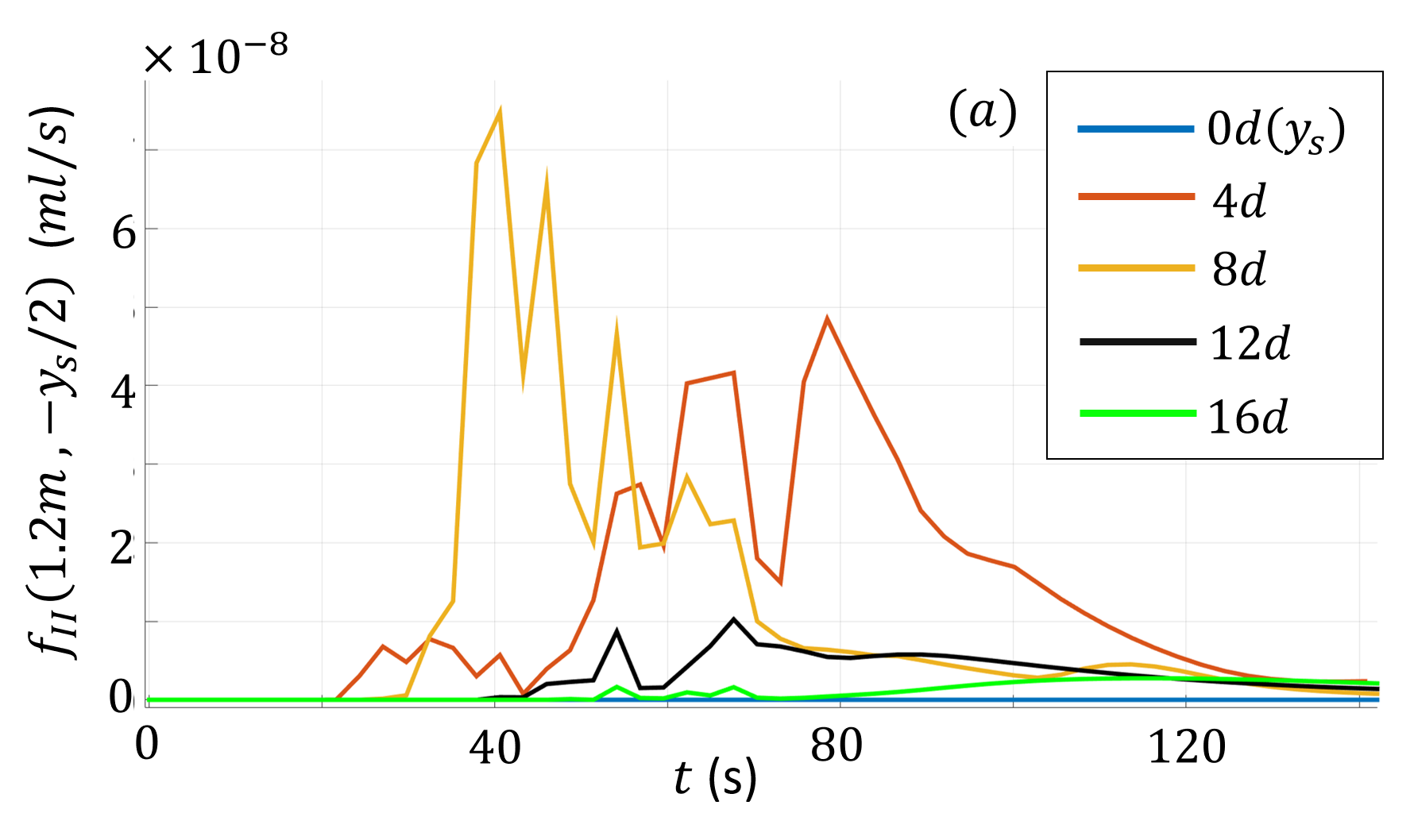}
\end{minipage}
\hspace{0.05\textwidth}%
\begin{minipage}{0.47\textwidth}\centering
  \includegraphics[width=1.0\textwidth]{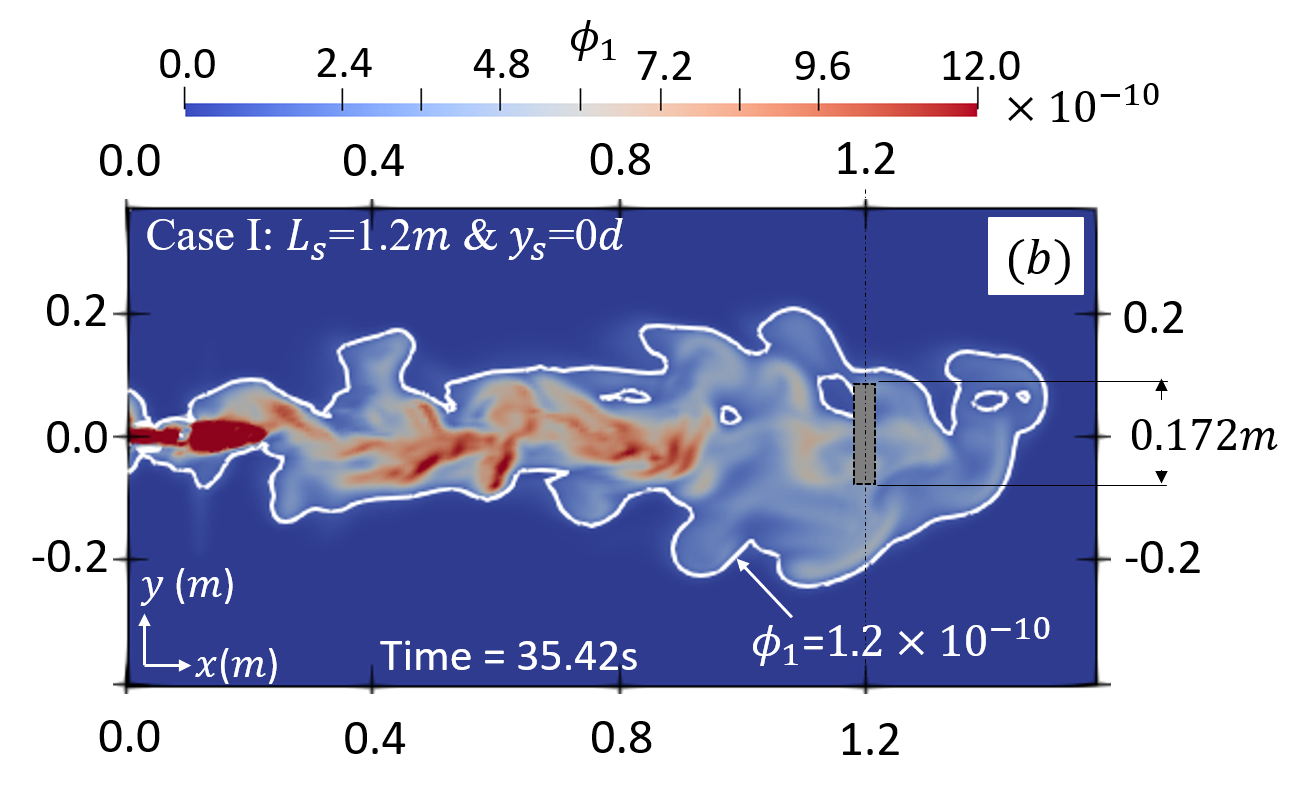}
\end{minipage}%
\begin{minipage}{0.47\textwidth}
  \includegraphics[width=1.0\textwidth]{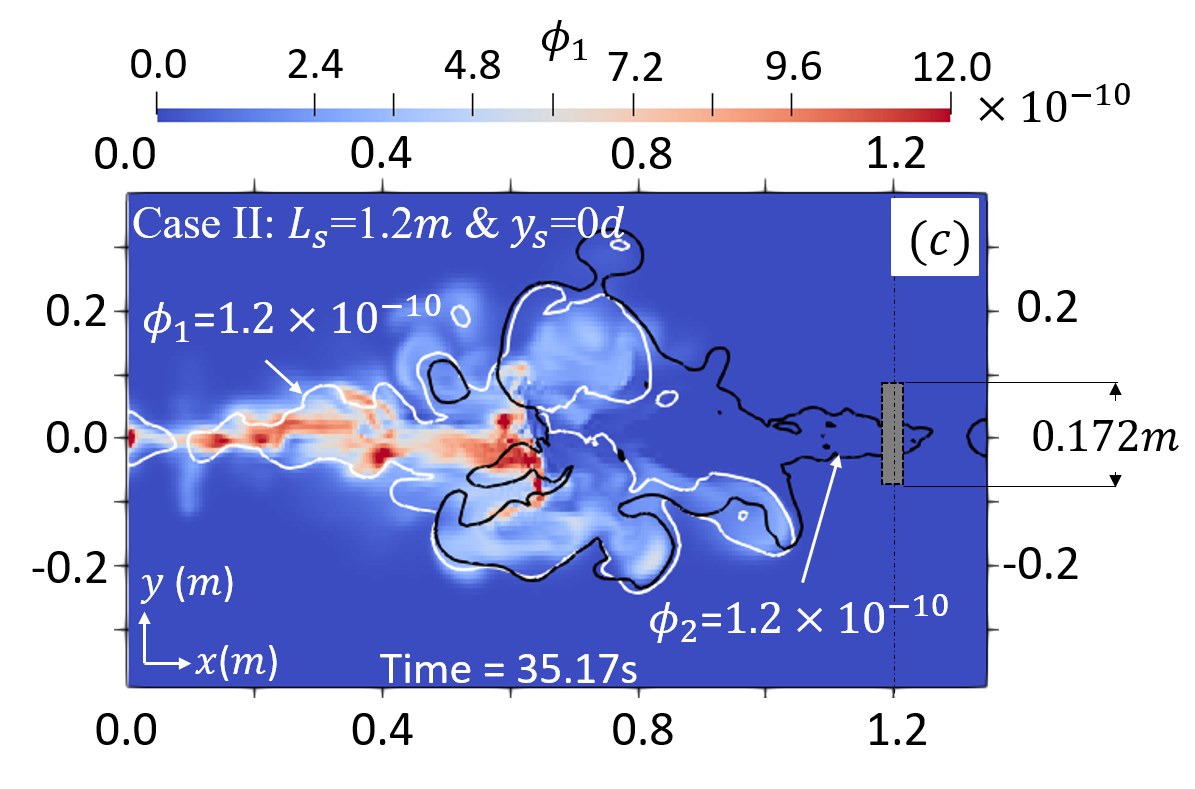}
\end{minipage}\\
\hspace{0.00\textwidth}%
\begin{minipage}{0.47\textwidth}
  \includegraphics[width=1.0\textwidth]{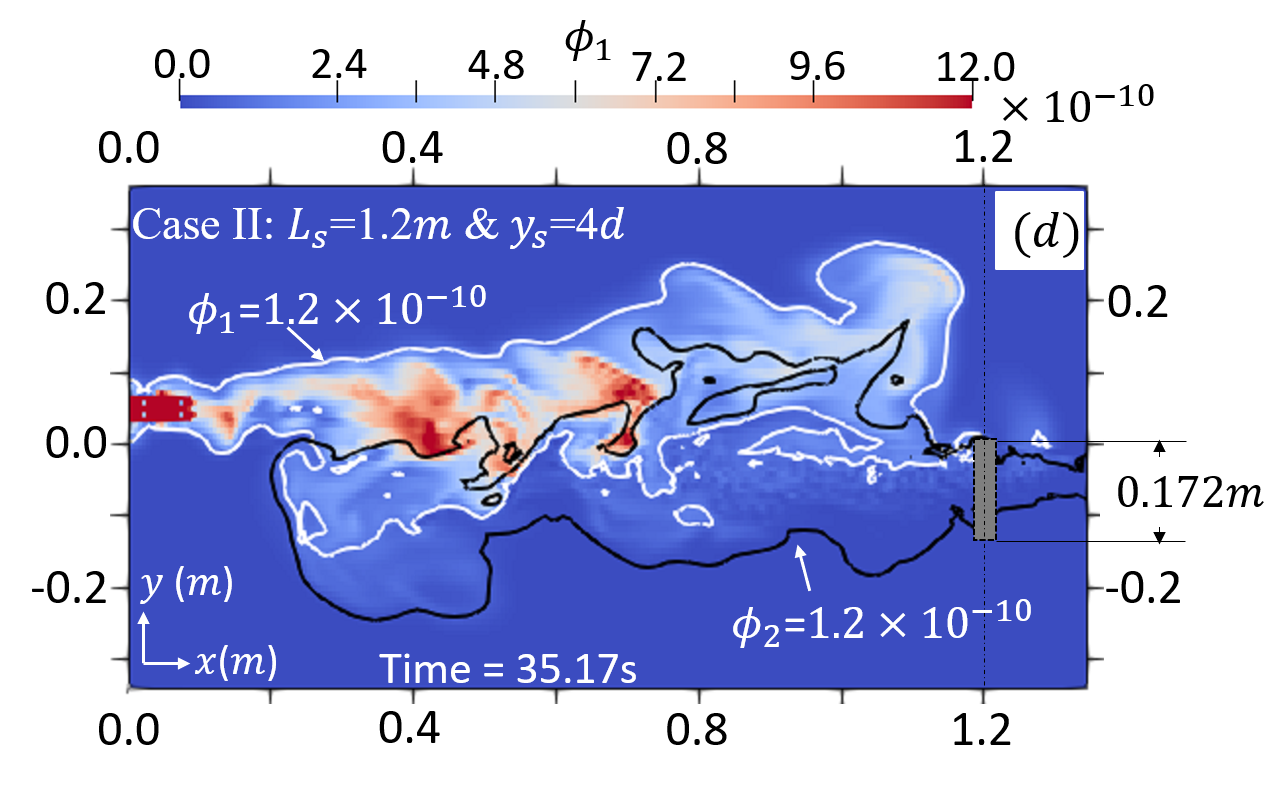}
\end{minipage}%
\begin{minipage}{0.47\textwidth}\centering
  \includegraphics[width=1.0\textwidth]{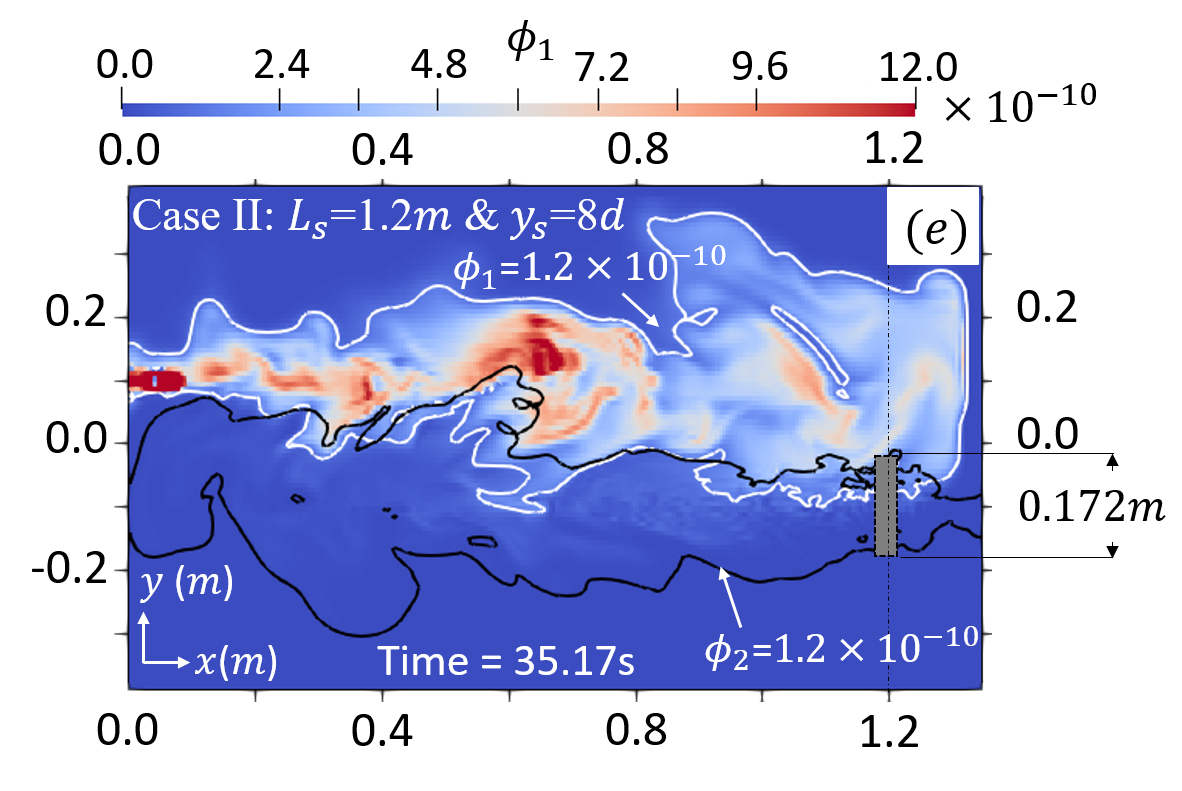}
\end{minipage}

\caption{Aerosol flux distribution and concentration contours for Case II (two-way conversation) compared with Case I (Person 2 silent). (a) Aerosol flux $f_{II}$ for $L_s=1.2m$ for different vertical separations $(y_s)$ for Case II. (b-e): Contours of aerosol concentration for $L_s=1.2m$ at $t \approx 35s$ for $y_s$ of (b) $0d$ in Case I, (c) $0d$ in Case II, (d) $4d$ in Case II and (e) $8d$ in Case II. In contrast to Case I shown in (b), it is evident in (c-e) that the passage of aerosol from one person to another is inhibited by the existence of two speech jets. The outlines of the jets are shown in white for Person 1 and black for Person 2. Filled colour contours are shown only for $\phi_1$. A side view of the circular region of interest (ROI) in front of Person 2 is represented by a grey rectangle. Here $\phi_1=\phi_o$  $(C_{s1}/C_{so})$ and $\phi_2=\phi_o  (C_{s2}/C_{so})$, (see supplementary movie S1 for the time evolution of the interaction between two speech jets for c-e). \label{fig4}}
\end{figure}

Thus far, the listener was entirely passive. In this set of simulations (Case II), Person 2 is present at a location $L_x=L_s+\Delta$ and $-y_s$ with respect to Person 1, i.e., Person 2 is of the same height or shorter than Person 1 (\hyperref[fig1]{Figure~\ref{fig1}b}). 
We calculate the flux of aerosols emanating from the mouth of Person 1 $(\phi_1)$ at the ROI located
in front of Person 2 (who is a susceptible individual) and denote it as $f_{II}(L_s,-y_{s}⁄2)$ (\ref{eq1}).
Simulations are performed for three different values of $L_s \in \{0.6\rm{m} ,1.2\rm{m} ,1.8\rm{m}\}$ with $y_s$ taking one of the seven values among $0d, 2d, 4d, 6d, 8d, 12d$ and $16d$ for each $L_s$; a couple of other values of $y_s$ are also chosen where needed.
As described earlier, each person speaks for 10 speech cycles in a staggered manner to have a total speech time of 70s (including silent intervals; \hyperref[fig1]{Figure~\ref{fig1}d}). The simulations are run longer than this duration, until almost all of the aerosols expelled by Person 1 pass through the ROI. \hyperref[fig4]{Figure~\ref{fig4}a} shows the time variation of $f_{II}$ for $L_s =1.2$m as $y_s$ is varied, which exhibits a non-monotonic behaviour. As $y_s$ increases from $0$, $f_{II}$ increases many folds until $y_s=8d$ and the time at which $f_{II}$ peaks (at a given $y_s$) goes on decreasing. As $y_s$ increases beyond $8d$, $f_{II}$ shows a decrease whereas the time at which it peaks increases (\hyperref[fig4]{Figure~\ref{fig4}a}).

\hyperref[fig4]{Figure~\ref{fig4}(c to e)} show the contour plots of the instantaneous aerosol concentration for Case II, for $y_s=0d,4d,8d$ respectively, with the color contours representing different values of $\phi_1$ and the line contours representing an iso-line for $\phi_1$ and $\phi_2$ equal to $1.2\times10^{-10}$; \hyperref[fig4]{Figure~\ref{fig4}b} shows the aerosol contours for Case I for comparison.
For $y_s =0d$ (Case II; \hyperref[fig4]{Figure~\ref{fig4}c}), the jets issuing out of the two orifices impinge on each other and “cancel” out each other’s effect to form a cloud in the middle. As a result, the flux of $\phi_1$ at Person 2 is much smaller for $y_s =0d$ than what it would be if Person 2 was silent as in Case I (see \hyperref[fig4]{Figure~\ref{fig4}b}). For $y_s=4d$, the jets are found to be “sliding over” each other with a region of overlap in the aerosol distributions (\hyperref[fig4]{Figure~\ref{fig4}d}). Although the head of flow from Person 1 could be seen deviating away from the Person 2, the overlapped region of $\phi_1$ manages to project a part of the aerosol flux on the latter’s face (grey rectangle). For $y_s  = 8d$ in \hyperref[fig4]{Figure~\ref{fig4}e}, the interference between the two jets is significantly reduced and therefore the aerosols from Person 1 find it easier to reach the ROI positioned in front of Person 2.
For higher $y_s$, only a small fraction of the speech jet from Person 1 can be expected to intersect the ROI, due to the large vertical separation. Thus, the competing effects of the jet interference and vertical separation lead to the observed maximum in the aerosol flux for an intermediate $y_s$ (\hyperref[fig4]{Figure~\ref{fig4}a}). The time evolution of the interaction of speech jets from Persons 1 and 2 for the cases shown in \hyperref[fig4]{Figure~\ref{fig4}c-e} has been presented as supplementary movie S1 ({see SI}). The values for $F_{II} (L_s,y_s,t_c )$ (\ref{eq2}) are calculated for all the simulations of Case II and used to calculate the virion exposure, as explained in next section.

\subsection{Viral exposure and infection probability for Cases I and II}

\begin{figure}[!h]
\centering%
\begin{minipage}{0.49\textwidth}\centering
  \includegraphics[width=1.0\textwidth]{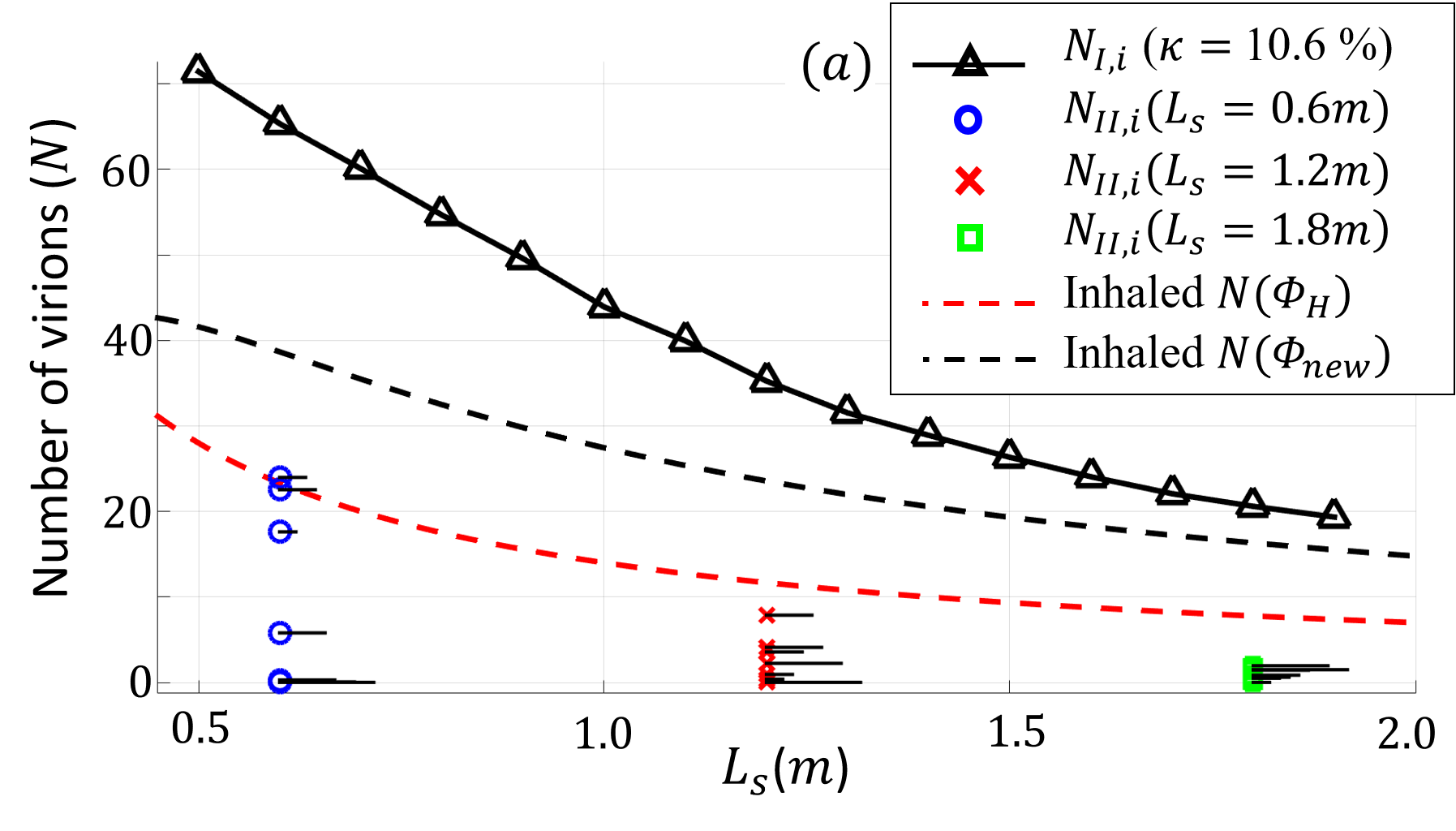}
\end{minipage}%
\begin{minipage}{0.49\textwidth}
  \includegraphics[width=1.0\textwidth]{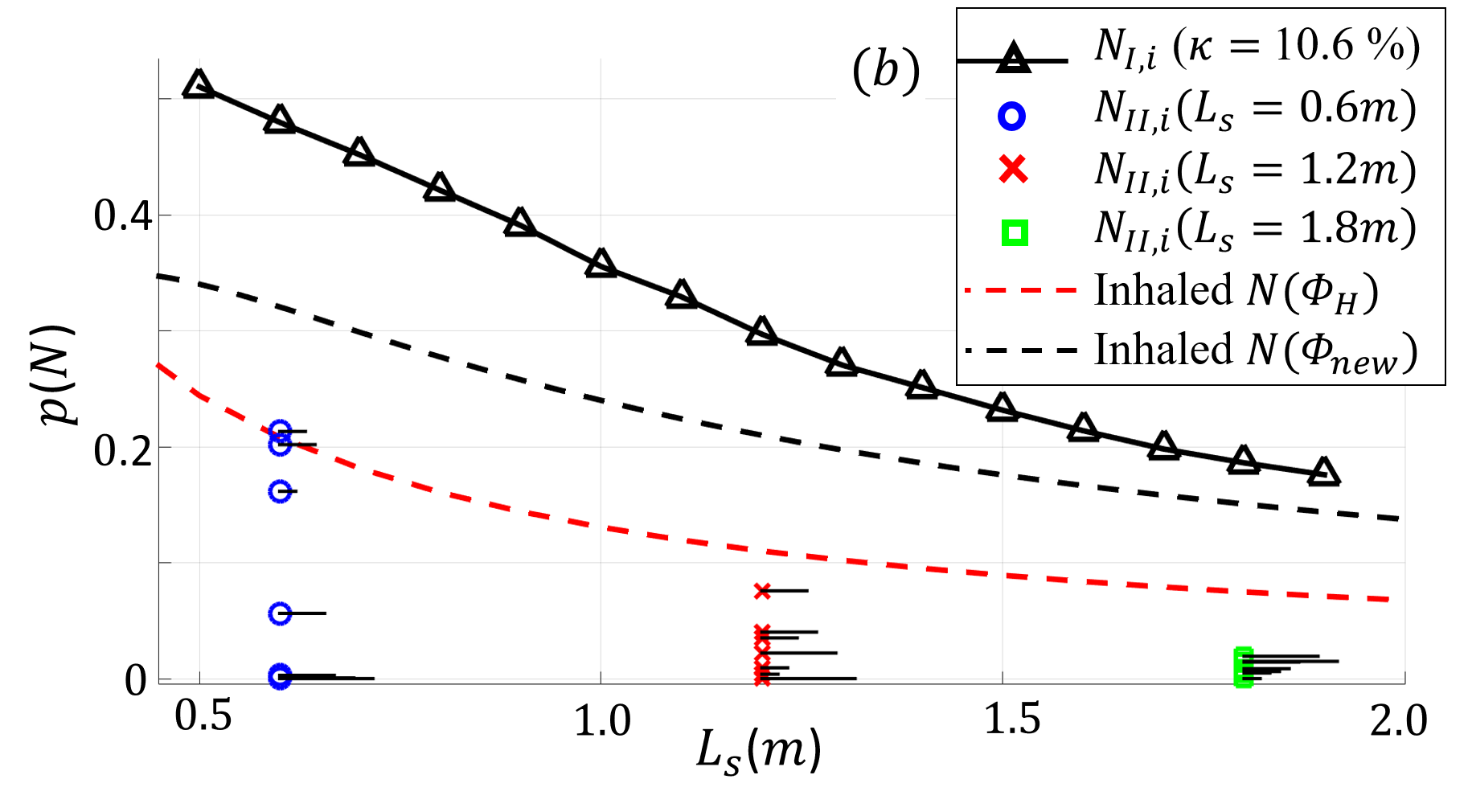}
\end{minipage}%
\caption { (a) The number of virions and (b) the probability of infection for Case I (curves) and Case II (symbols). Each symbol for Case II corresponds to a different simulation. The probability of infection in Case II is always much lower than in Case I, which means that a dialogue is always better than a monologue. The red-dashed curve, showing viral exposure through inhalation alone (Yang et al.,2020), is always lower by about 50\% than the solid black curve which includes exposure to the eyes and the mouth. The locations of the symbols correspond to the axial separation $L_s$ between the speakers and each symbol is accompanied by a horizontal bar whose length is proportional to $y_s$, for $y_s \in \{0d,2d,4d,6d,8d,12d,16d\}$. In Case II, the infection probability for a given $L_s$ varies non-monotonically with $y_s$.
}
\label{fig5}
\end{figure}

In this section, we determine the virion ingestion by a susceptible individual due to the aerosol transport from speech flows. First, the total virion exposure to a passive listener in Case I is calculated using inhaled virions as well as the ones projected onto the eyes and mouth area. The number of inhaled virions in Case I is calculated using the same expression as in (\ref{eq3}) but by replacing $\Phi_H$ by $\Phi_{new}$ (\ref{eq4}), as

\begin{equation}
    N(\Phi_{new} )=\ c_v \Phi_{new} Q_r\ t \label{eq5} 	 
\end{equation}
We use $c_v=7\times10^6 \rm{ml}^{-1}$ (Yang et al., \hyperlink{bib49}{2020}). To calculate the infection through eyes and mouth, we first obtain the total number of virions projected onto the listener’s face (or passing through the ROI in front of their face) as $N_{I,f}=c_v F_I (L_s,y_s,140\rm{s})$, i.e., integrated over 140s. Towards this, we use $F_I$ for $y_s=0$ (blue curve in \hyperref[fig2]{Figure~\ref{fig2}b}) which corresponds to the maximum aerosol exposure for Case I. We assume that viral exposure to the eyes and mouth is the fraction $\kappa$  of $N_{I,f}$, where $\kappa=A_{em}/A_{ROI}$; here $A_{em}$ is the total area of eyes and mouth, and $A_{ROI}$ is the area of the ROI (i.e., the projected face area). The total number of virions that are ingested by the listener (potentially causing infection) for Case I is, therefore,
\begin{equation}
    N_{I,i}\ =\ N(\Phi_{new} )\ +\ \kappa \cdot N_{I,f}\ \label{eq6} .			
\end{equation}
Reasonable estimates for the areas of the eyes (radius $\approx 1.2\rm{cm}$) and of the mouth and lips area ($ \approx 15.55 \rm{cm}^2$), suggest that these amount to about $10.6\%$ of the area of the face $(\kappa=0.106)$. 

For Case II the total exposure to Person 2 over 140s from the aerosols expelled by Person 1 is denoted by $F_{II} (L_s,-0.5*y_s,140\rm{s})$ (\ref{eq1} and \ref{eq2}). The total virion exposure to the face of Person 2 is therefore given as $N_{II,f}\ =\ c_v F_{II} (L_s,-0.5*y_s,140\rm{s})$. To calculate the virion exposure causing infection we assume that 
\begin{equation}
   N_{II,i}/N_{II,f} \ =\ N_{I,i}/N_{I,f}  , \label{eq7} 
\end{equation}
where $N_{II,i}$ represents the number of virions ingested by Person 2 through inhalation and exposure to eyes and mouth for a given simulation of Case II. Note that $N_{I,i}$ and $N_{II,i}$ represent conservative estimates of ingested virions, assuming that the entire virion exposure to eyes and mouth leads to infection.

\hyperref[fig5]{Figure~\ref{fig5}a} presents a summary plot of the virion exposure to Person 2 (for Cases I and II) obtained using different measures. The number of inhaled virions using the method of Yang et al. (\hyperlink{bib49}{2020}), $N(\Phi_H)$(\ref{eq3}), is plotted as a red dashed curve, whereas $N(\Phi_{new})$ obtained by averaging over a localized region (\ref{eq5}) is plotted as a black dashed curve. Note that the region $x<0.45$m is not considered as self-similarity of speech flow does not hold and calculations using steady state parameters are expected to be unrealistic in this region (Fig. S5b). As can be seen from \hyperref[fig5]{Figure~\ref{fig5}a}, $N(\Phi_{new})$ is nearly twice that of $N(\Phi_H)$ for the entire range of separation distances, $L_s$, consistent with \hyperref[fig3]{Figure~\ref{fig3}} . Note that both $N(\Phi_H)$ and $N(\Phi_{new})$ have been calculated in the context of Case I, i.e., with Person 2 as a passive listener. 
The total number of ingested virions for Case I, $N_{I,i}$, shows even higher values as compared to $N(\Phi_{new})$ as expected (\ref{eq6}). 
We propose that $N_{I,i}$ is a more realistic (although conservative) measure for the viral load for determining the risk of infection, as compared to the estimate based on inhaled virions alone. The number of ingested virions for Case II, $N_{II,i}$,
is shown as symbols for three different values of $L_s$. The maximum number of ingested virions for this case at each $L_s$ is seen to occur at an intermediate value of $y_s$ consistent with \hyperref[fig4]{Figure~\ref{fig4}a}. \hyperref[fig5]{Figure~\ref{fig5}a} clearly shows that the viral exposure to Person 2 when they are actively engaged in a conversation (Case II) is lower as compared to that when Person 2 is a passive listener (Case I) by a factor more than two. This is due to the interference of the speech jets from the two persons during a conversation as discussed earlier. 

\begin{figure}[!h]
\centering%
\begin{minipage}{0.44\textwidth}\centering
  \includegraphics[width=1.0\textwidth]{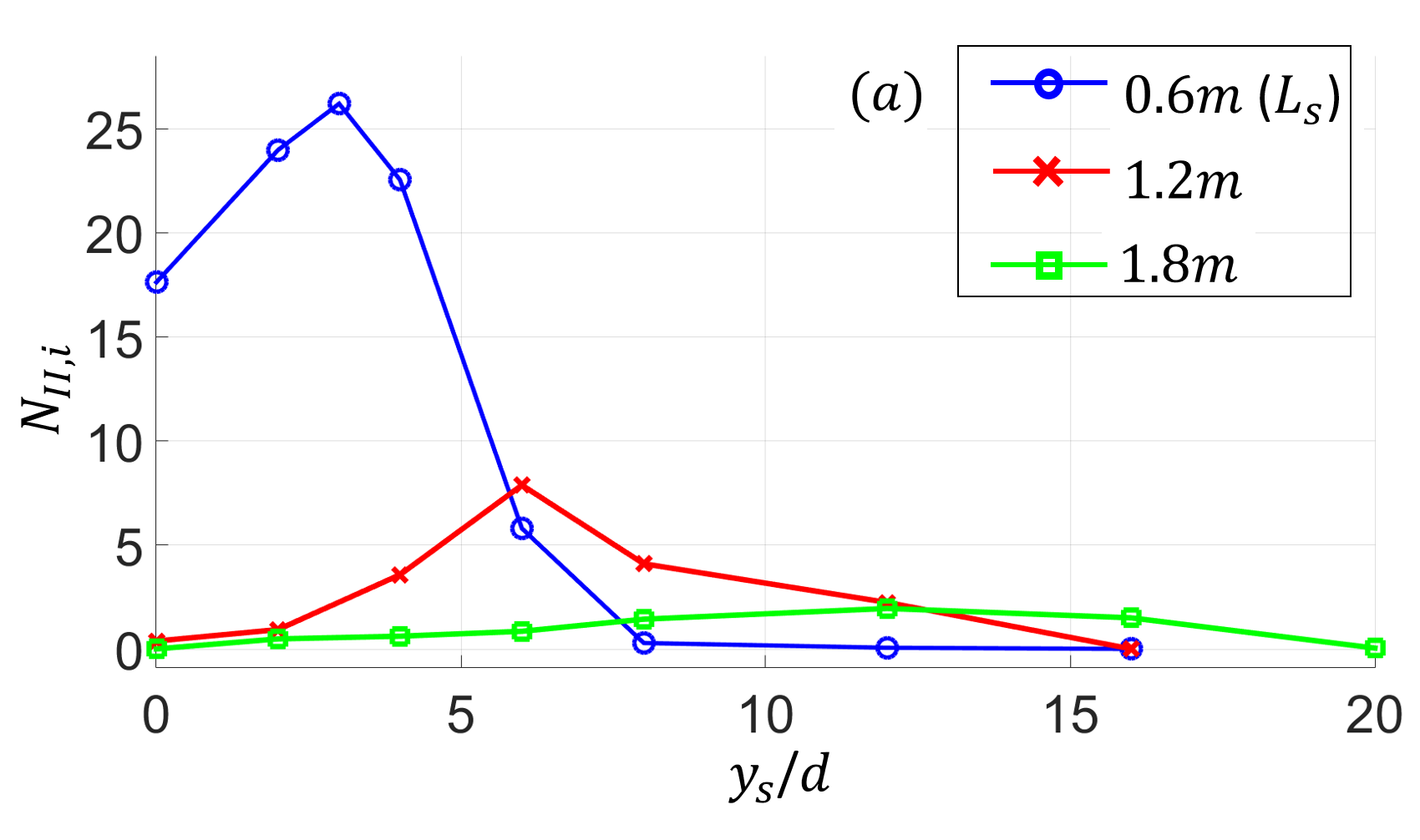}
\end{minipage}%
\hspace{0.08\textwidth}%
\begin{minipage}{0.42\textwidth}
  \includegraphics[width=1.0\textwidth]{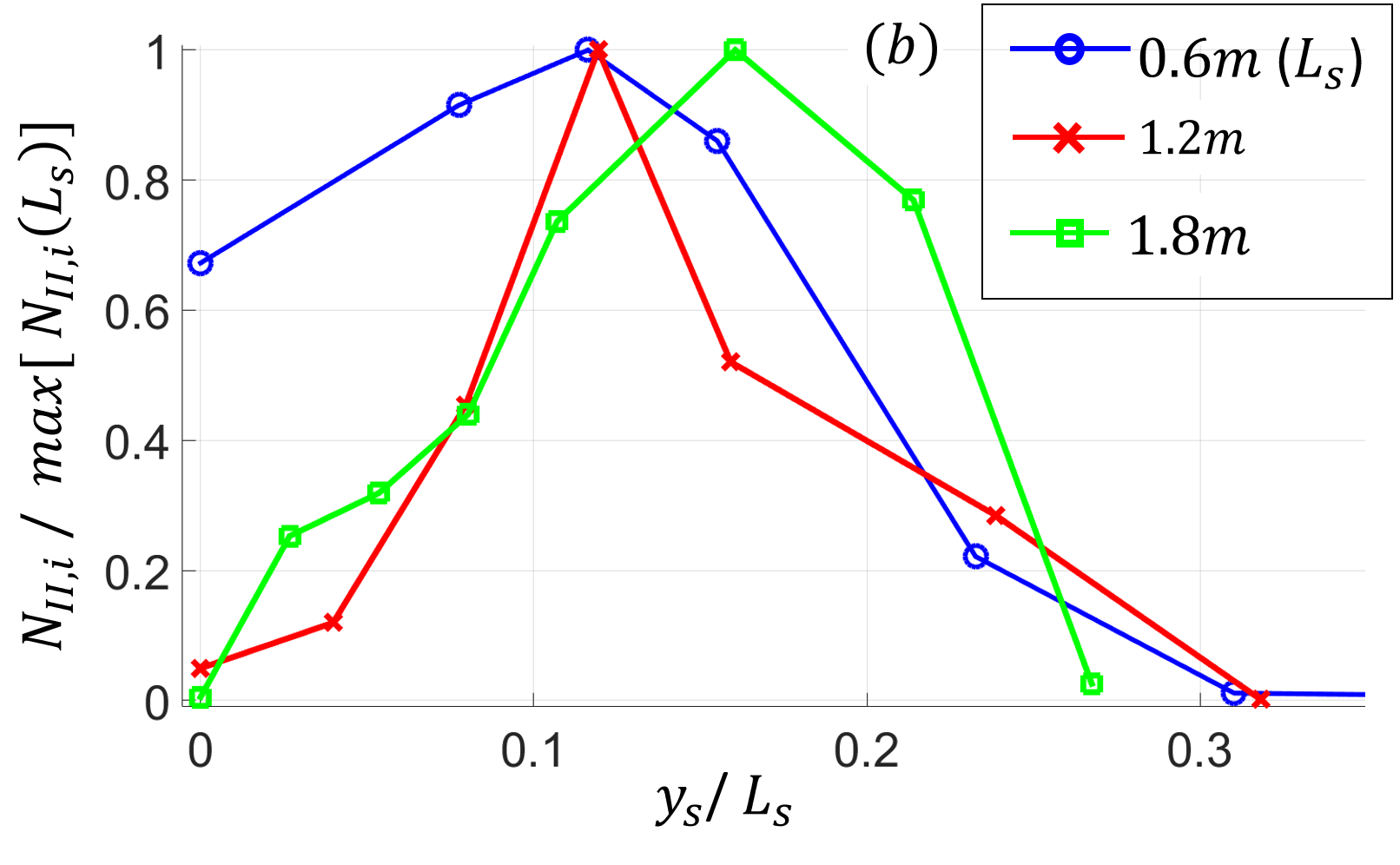}
\end{minipage}%
\caption{ (a) The number of virions $N_{II,i}$ ingested by Person 2 as a function of the vertical separation $y_s/d$ for different axial separations $L_s$. (b) Variation of normalized $N_{II,i}$ with $y_s/L_s$. Risk of infection varies non-monotonically with vertical separation.  \label{fig6}}
\end{figure}

In \hyperref[fig5]{Figure~\ref{fig5}b} we plot the probability of infection corresponding to the viral load $N$ calculated as  
\begin{equation}
    p(N)=1-exp(-N/N_{inf}),\label{eq8}
\end{equation}
where $N_{inf}\approx 100$ is the characteristic number of virions causing infection (Yang et. al., \hyperlink{bib49}{2020}). For Case I, $p(N_{I,i} )$ is found to be as high as $0.5$ when $L_s=0.6$m. Thus, there is a real risk of inflection to a passive listener for a 2ft separation distance, even when the other person speaks for a very short time (speech time =70s and exposure time = 140s). As the separation distance increases the infection probability decreases monotonically, with  $p(N_{I,i} )<0.2$ for $L_s=1.8$m (\hyperref[fig5]{Figure~\ref{fig5}b}). Thus, it is much safer to adhere to the 6ft rule even while listening to someone speak for a short span of time, in order to minimize the risk of infection. For Case II involving conversations, $p(N_{II,i}$ ) is much less than $p(N_{I,i} )$ as expected and the risk of infection is therefore low. The maximum $p(N_{II,i} )$ among all the conversation cases is about $0.2$ for $L_s=0.6$m for a conversation of a little over a minute. For a longer conversation or for asymmetric speech times (see the next section) between the people, the risk of infection can be expected to be higher than this. Increasing separation between people during conversation is again a sure-shot way of reducing the infection probability (\hyperref[fig5]{Figure~\ref{fig5}b}).

\hyperref[fig6]{Figure~\ref{fig6}a} provides a graphical representation of the dependence of $N_{II,i}$ on $L_s$ and $y_s$. As follows from the discussion above, the viral dose to Person 2 decreases considerably with increase in $L_s$ and $N_{II,i}$ peaks at an intermediate vertical separation $y_s$ (see also \hyperref[fig5]{Figure~\ref{fig5}}). \hyperref[fig6]{Figure~\ref{fig6}b} plots the variation of the normalized viral dose (by its peak value for a given $L_s$) with $y_s/L_s$. We find that maximum of the normalized $N_{II,i}$ occurs at $y_s/L_s$ of $0.12-0.15$ which can be a useful result from the scaling point of view. Furthermore, $N_{II,i}/max(N_{II,i})$ is significantly reduced for $y_s⁄L_s > 0.3$, implying that for a given separation distance between two people, the viral load during a short conversation can be expected to be low if the condition $y_s \gtrsim L_s⁄3$ is satisfied.

\subsection{Case IIt: Temporal asymmetry in speech during conversation}
\begin{figure}[!h]
\centering%
\begin{minipage}{0.61\textwidth}\centering
  \includegraphics[width=1.0\textwidth]{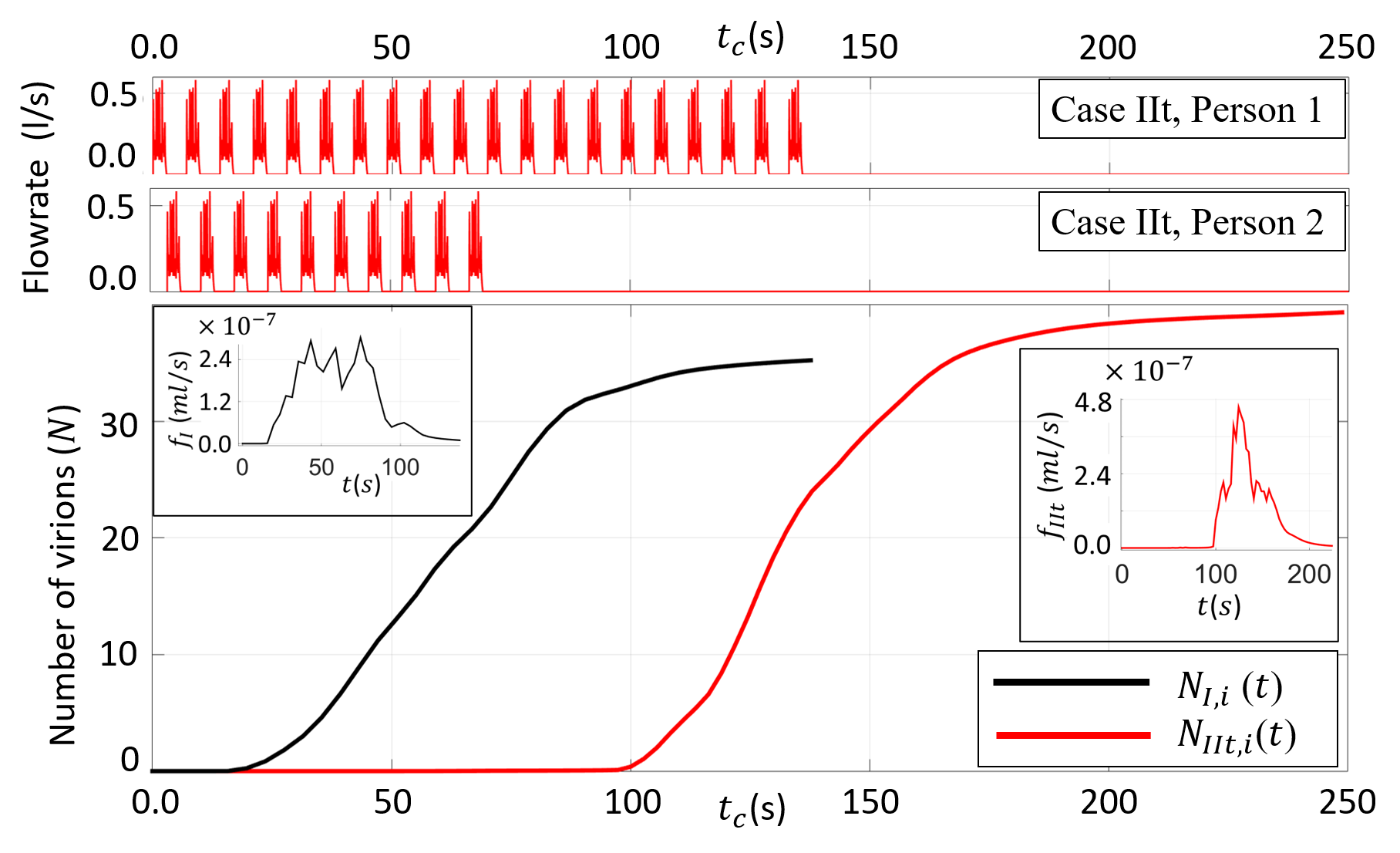}
\end{minipage}%
\caption{The role of temporal asymmetry in determining viral exposure, shown by plotting (bottom panel) the number of virions ingested as a function of time in Case IIt compared to Case I. The panels at the top show the inlet flow rates. For both cases $L_s=1.2m$ and $y_s=0$. For Case IIt, the conversation does not stop incoming aerosols from reaching Person 2 but merely delays it by about 100s (see supplementary movie S2 for time evolution of the flow for this case). The time variation of $N_{IIt,i}$ shown here can in principle be extrapolated to estimate the viral load for conversations longer than used here.  \label{fig7}}
\end{figure}

So far, we have presented conversation cases where the number of speech cycles and the total speech time were the same for Persons 1 and 2 (\hyperref[fig1]{Figure~\ref{fig1}d}). This corresponds to a scenario wherein the exchange of short phrases during a conversation is more or less “symmetric” (although staggered in time). The objective of this exercise was to quantify the difference between one-way and two-way conversations. However, in reality, conversations can take place in a variety of different ways and in most of the cases there is likely to be an asymmetry of speech between two people. Here we present one instance of an “asymmetric” conversation (Case IIt) wherein both the persons engage in a symmetric conversation for 70s, after which Person 2 stops talking while Person 1 continues for another 70s; see \hyperref[fig7]{Figure~\ref{fig7}}. Both the persons are considered to be of equal height $(y_s=0)$ with a separation distance, $L_s=1.2$m, and the simulation is run for about 250s. \hyperref[fig7]{Figure~\ref{fig7}} shows the variation of the number of virions ingested, $N_{IIt,i}$, with time interval, $t_c$, obtained by varying $t_c$ in the calculation of $F(L_s,y_s,t_c)$ in (\ref{eq2}) for Case IIt. Also shown for comparison is $N_{I,i} (t_c)$ for Case I representing a one-way conversation. It is seen that the number of ingested virions for Case IIt is negligibly small until about $100$s (cf. \hyperref[fig4]{Figure~\ref{fig4}a}) after which $N_{IIt,i}$  starts increasing as the aerosols from Person 1 (who continues to speak after 70s) reach Person 2. The increase in $N_{IIt,i}$ is initially rapid but tapers off after Person 1 stops speaking after 140s. Overall, the shapes of curves and the number of virions ingested by Person 2 at the end of conversation are similar between $N_{IIt,i}$ and $N_{I,i}$ (\hyperref[fig7]{Figure~\ref{fig7}}).

The probability of infection for Person 2 at $t_c=250$s for Case IIt comes out to be $p(N_{IIt,i}) = 0.32$ (\ref{eq8}), which is slightly higher than that obtained for Case I (\hyperref[fig5]{Figure~\ref{fig5}b}). This suggests that whenever there is an asymmetry in speech between people engaged in a conversation, the person who talks less is at an enhanced risk of catching infection than the person who talks more. This is true even when both the persons are of nearly the same height. The time evolution of the speech jets for Case IIt is presented as supplementary movie S2 ({see SI}). Note that the asymmetry in time need not occur at the end of a conversation but could be embedded within it, in which case the curve for $N_{IIt,i}$ in \hyperref[fig7]{Figure~\ref{fig7}} will show an “up and down” variation, providing temporally varying and non-monotonic probability of infection. 

\section{Discussion}

\subsection{Lateral spread and temporal evolution of the interacting speech jets}
\begin{figure}[!h]
\centering%
\begin{minipage}{0.49\textwidth}\centering
  \includegraphics[width=1.0\textwidth]{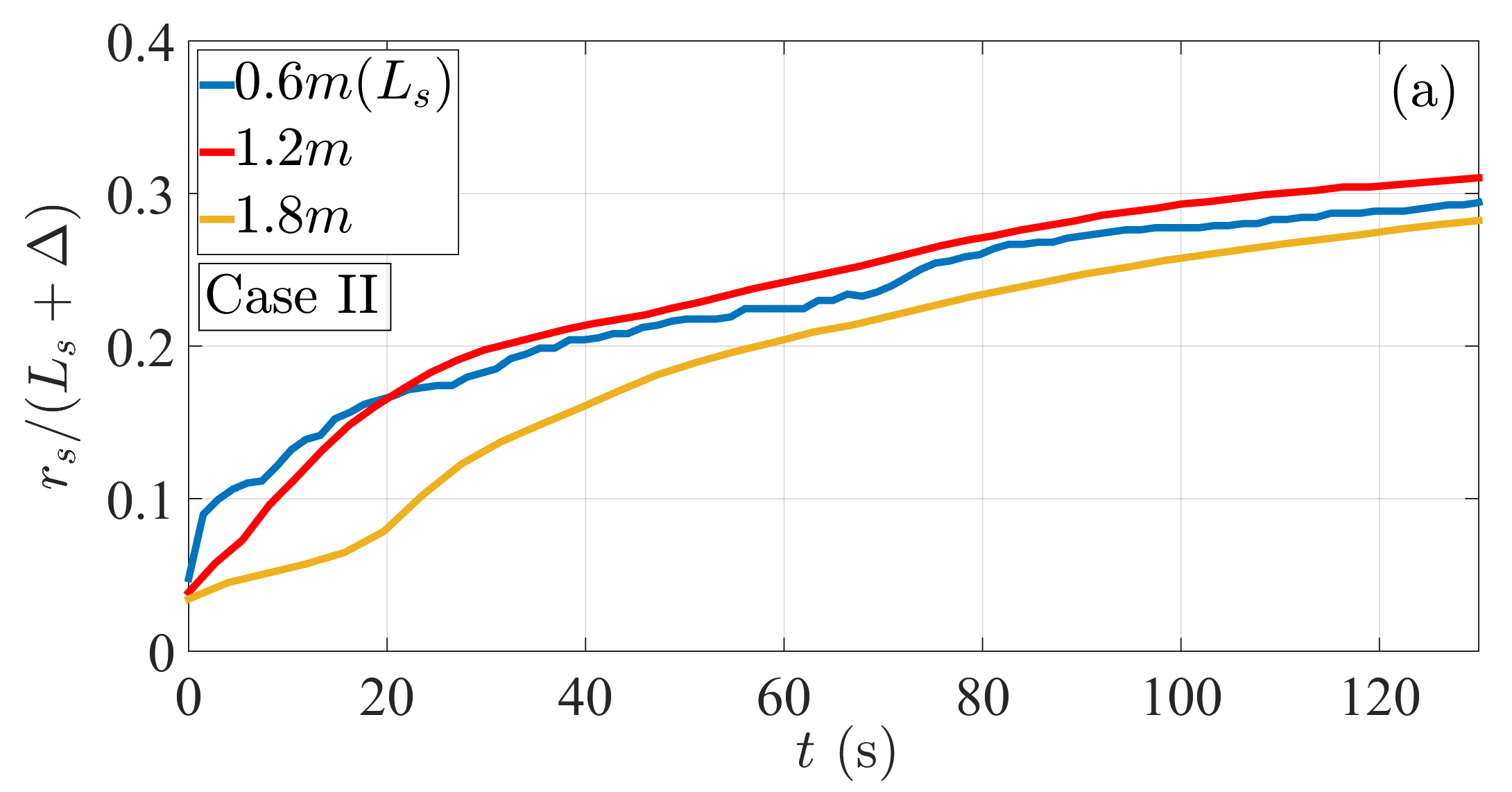}
\end{minipage}
\begin{minipage}{0.48\textwidth}
  \includegraphics[width=1.0\textwidth]{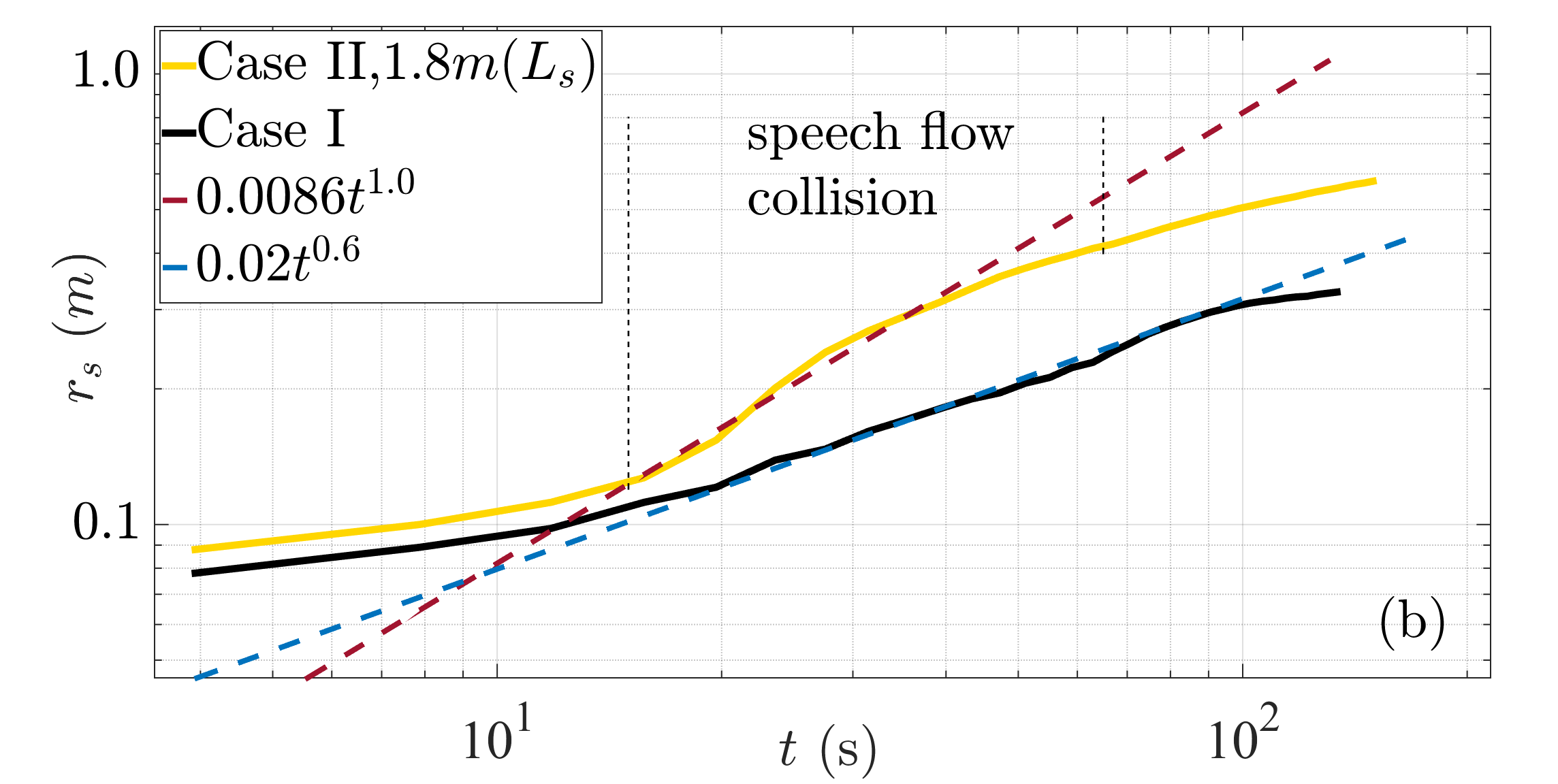}
\end{minipage}%
\caption{\label{fig:lateral_width}(a) Lateral width ($r_s$ from (\ref{eq9})) normalised by the corresponding domain length between the two persons for Case II. (b) Lateral width of the flow on a log-log scale for Cases I and II, with $L_s=1.8m$. For Case II, $y_s = 0$.}
\end{figure}

Although the collision of the speech jets during conversations can restrict the axial transport of aerosols (especially for people of comparable heights; \hyperref[fig4]{Figure~\ref{fig4}}), it can result in considerable lateral spread of the flow. This aspect is relevant to how the interaction of two people generates a ``cloud'' of infected aerosols that might affect disease spread in poorly ventilated rooms. We define the lateral width ($r_s$) of the flow using
\begin{equation}
\int_{r=0}^{r_s} \int_{\theta} \int_x \phi r dr d\theta dx =0.9 \int_{r=0}^{\infty} \int_{\theta} \int_x \phi r dr d\theta dx.  \label{eq9}
\end{equation}
The lateral widths of the flow in Case II (two-speaker) simulations are compared for different axial separations in \hyperref[fig5]{Figure~\ref{fig:lateral_width}a}, and with   Case I (single-speaker case) in \hyperref[fig5]{Figure~\ref{fig:lateral_width}b}. The axial length of the domain, $L_s +\Delta$ ($\equiv L_x$; see \hyperref[fig1]{Figure~\ref{fig1}b}), acts as a rough scaling length for the lateral spread \hyperref[fig:lateral_width]{(Figure~\ref{fig:lateral_width}a)}; $r_s/(L_s+\Delta)$ reaches a value of 0.3 after $120$s, which is about $0.55$m for $L_s=1.8$m. \hyperref[fig:lateral_width]{Figure~\ref{fig:lateral_width}b} shows that the collision of the speech jets from the two persons initially increases the rate of lateral spread, ($r_s\propto t^{1.0}$), which is significantly higher than that in Case I where lateral spread is $\propto t^{0.6}$. This provides an estimate of the rate of evolution of the aerosol cloud that might affect a third person positioned nearby; it should also be possible to extrapolate the rate of growth for longer conversation times using the power laws in \hyperref[fig:lateral_width]{Figure~\ref{fig:lateral_width}b}. At later times, i.e. after the conversation ends, the rate of spread in Case II becomes slower, but the lateral spread continues to remain much larger than that for Case I.
\begin{figure}[!h]
\centering%
\begin{minipage}{0.49\textwidth}
  \includegraphics[width=1.0\textwidth]{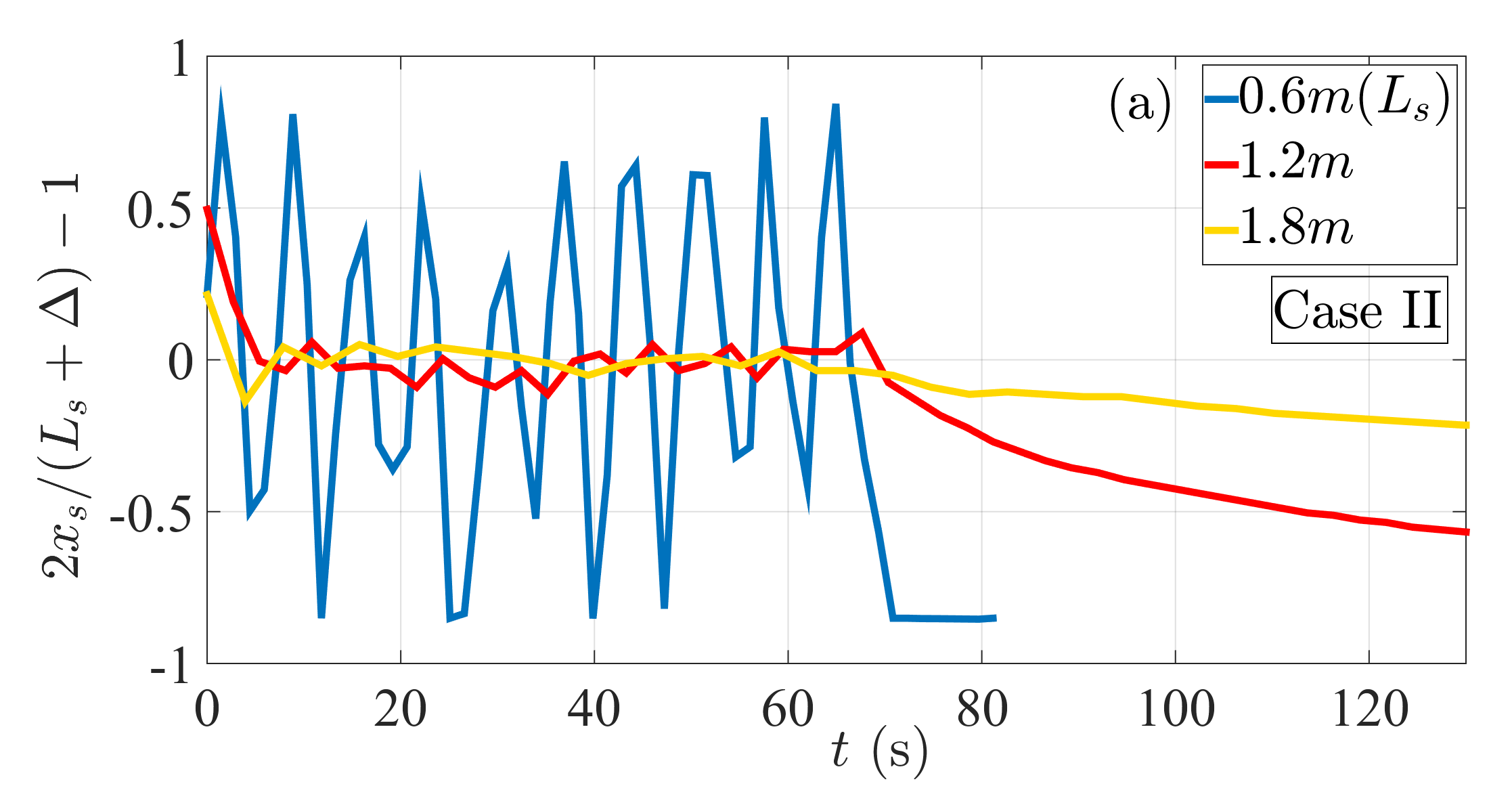}
\end{minipage}%
\begin{minipage}{0.49\textwidth}\centering
  \includegraphics[width=1.0\textwidth]{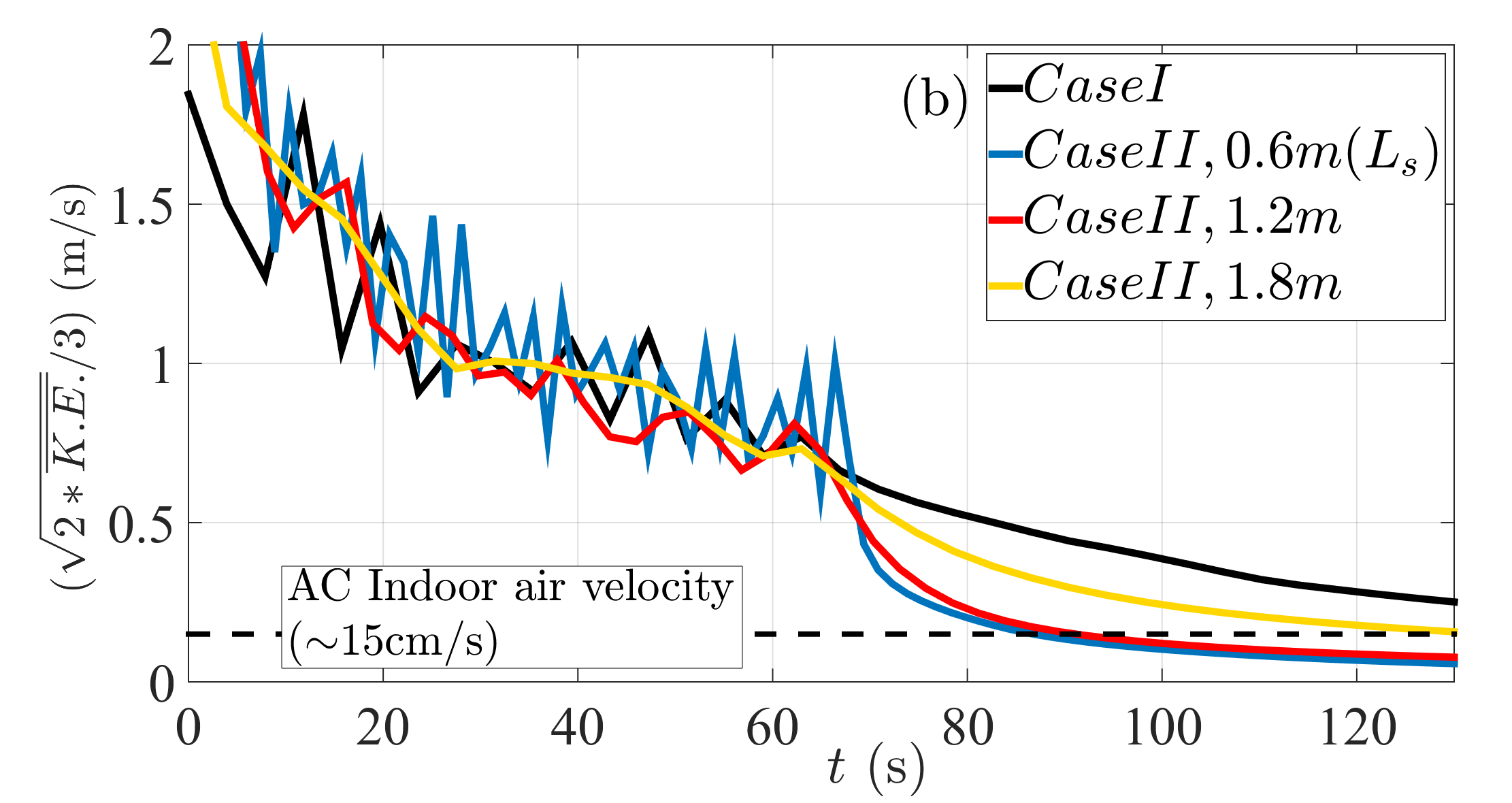}
\end{minipage}
\caption{\label{fig:KE_oscillation}(a) Flow oscillation due to the collision of alternate speech jets from two people, represented by the stagnation point ($x_s$) in the centreline velocity profile. The location $x_s$ is normalized such that Person 1 is at $x=-1$ and Person 2 is at $x=1$. (b) A measure of average velocity, based on the domain-averaged kinetic energy, as a function of time for Cases I and II. For Case II, $y_s=0$.} 
\end{figure}

In \hyperref[fig:KE_oscillation]{Figure~\ref{fig:KE_oscillation}a}, we plot the location of the stagnation point, $x_s$, between the two speech jets for Case II. For small axial separation as $L_s=0.6$m, we see that the location of the stagnation point nearly reaches either speaker at $(2x_s/(L_s+\Delta))-1 =\pm1$. This wide oscillation is responsible for the non-zero aerosol flux even in a face-to-face ($y_s=0$) conversation for $L_s=0.6$m (see \hyperref[fig6]{Figure~\ref{fig6}}). This is not true, however, for larger axial separations $L_s$. The oscillation of the stagnation point is a feature of the pulsatile nature of the flow and is not captured if the flows are taken to be steady jets (for which the stagnation point would remain fixed in place at mid-domain). The oscillation of the stagnation point also suggests that the ``safe'' axial separation between two persons facing each other depends on the duration of the speech pulses in \hyperref[fig:KE_oscillation]{Figure~\ref{fig:KE_oscillation}a}. Following Abkarian et al., \hyperlink{bib1}{(2020)}, the time $(t^*)$ taken by a jet to reach the separation distance $L_s$ is given as $t^*=L_s^2 tan\alpha/ 2\bar{v}_od$, which gives $t^*$= (3.34s, 13.35s, and 30.04s) for $L_s$= (0.6m, 1.2m ,and 1.8m) respectively. For $L_s=0.6$m, $t^*=3.34$s is comparable to the half of the speech-cycle duration, which is $3.5$s. Note that the  average orifice velocity $\bar{v}_o = 0.39m/s$.

\hyperref[fig:KE_oscillation]{Figure~\ref{fig:KE_oscillation}b}, plots the time variation of a measure of velocity related to the averaged kinetic energy as $\sqrt{2\overline{K.E.}}/3$, with $\overline{K.E.}$ defined as follows. 
\begin{equation}
 \overline{K.E.}=\frac{1/2 \int (u^2+v^2+w^2) dV |_{\phi>0.01\phi_o}}{\int dV|_{\phi>0.01\phi_o}},  \label{eq10}
\end{equation}
 where $V$ represents the domain volume. The decay of the domain-averaged kinetic energy is nearly identical for Cases I and II until the end of the speech duration (i.e., 70s), after which the kinetic energy decays at different cases for different cases. The decay rate is larger for the two lower values of  $L_s$ (Case II), presumably because of the enhanced dissipation resulting from more intense gradients generated by the collision of speech jets; \hyperref[fig:KE_oscillation]{Figure~\ref{fig:KE_oscillation}b}. The residual velocity in the domain reaches fairly low values after the conversation ends and can soon become comparable to the speed of the background air motion found in typical indoor environments. For example, a threshold background current in an air-conditioned room is of the order of 15cm/s (\hyperlink{bib27a}{Matthews et al. 1989}), shown by dashed horizontal line in \hyperref[fig:KE_oscillation]{Figure~\ref{fig:KE_oscillation}b}. Under such conditions, the transport of virus-laden aerosols can be expected to switch to the airborne transmission mode.

\subsection{Relevance towards public health guidelines and epidemiological modelling}
\begin{figure}[!h]
\centering%
\begin{minipage}{0.44\textwidth}\centering
  \includegraphics[width=1.0\textwidth]{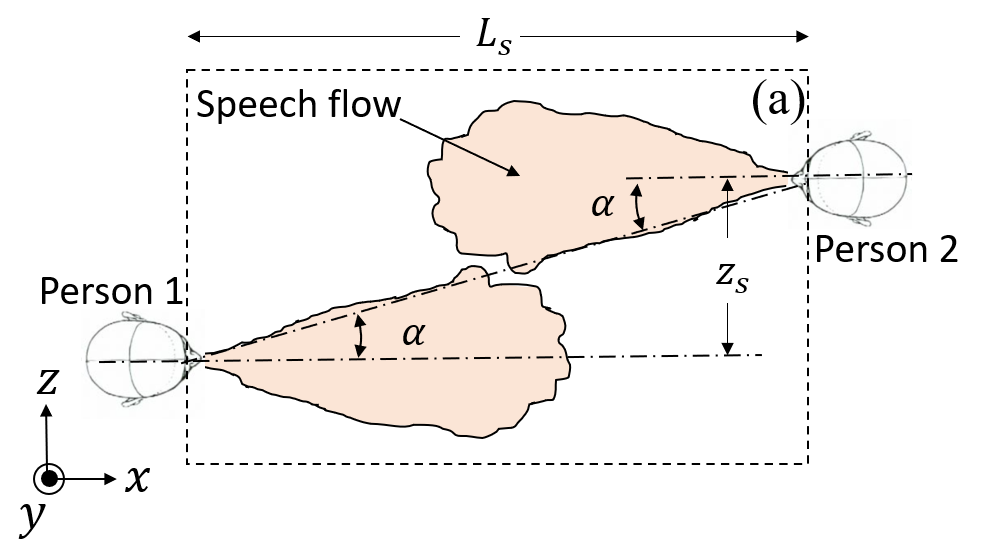}
\end{minipage}%
\begin{minipage}{0.44\textwidth}
  \includegraphics[width=1.0\textwidth]{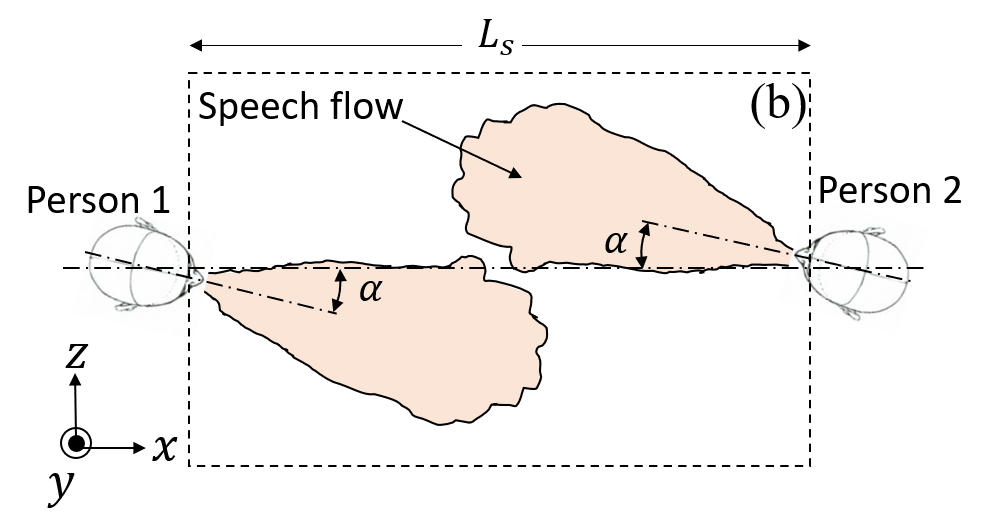}
\end{minipage}%
\caption{Lateral/angular separation to minimize transmission risk. (a) Schematic of two persons separated laterally by a distance $z_s$ such that the risk of infection is minimized. (b) Schematic of two people conversing with their heads turned away from each other by an “angular separation” to achieve the same effect as in (a). For an axial separation up to 6ft, a tilt angle $(\alpha)$ of $9^\circ$ or more is recommended.  \label{fig8}}
\end{figure}

Our results provide guidance, in addition to physical distancing, for safe interaction among people engaged in short conversations (up to about a minute or two). This situation arises routinely in day-to-day life, for example in shopping malls or stores, where over-the-counter conversations take place for short durations. We show the effect of buoyancy to be negligible, since the rate of spread of the speech flow in the vertical and horizontal planes is practically identical (see Fig. S3). So our results on the height difference (vertical separation $y_s$) between two people also apply  to  lateral separation (denoted by $z_s$; \hyperref[fig8]{Figure~\ref{fig8}}). This is an important result since the lateral separation between people can be controlled but the height difference cannot! We find surprisingly that the probability of infection from a two-way conversation is maximum, not when the two people face each other directly, but when  $y_s$ (or $z_s$) is about 12-15\% of $L_s$ (\hyperref[fig6]{Figure~\ref{fig6}b}); for an axial separation of 4ft, this amounts to about 15cm of vertical or lateral separation. However, infection probability is considerably lower when
$y_s⁄L_s$ (or $z_s⁄L_s$ ) 
is greater than 0.3 (\hyperref[fig6]{Figure~\ref{fig6}b}) and this helps in devising the following general rule for minimizing the risk of infection. When people in a conversation turn their faces away from each other very slightly, it can be worse than facing each other directly, but an angle greater than $tan^{-1}  (0.18/1.2)\approx 9^\circ$ (\hyperref[fig8]{Figure~\ref{fig8}b}) can be very helpful. This implies a total “angular separation” of $18^\circ$ or higher between two people for a safe conversation which still enables eye contact. This angular separation matches well with the angle of spread of the speech jet ($20^\circ$; Fig. S3), which means that the speech jet from one person effectively misses the face of the other person, thereby minimizing viral exposure. As expected, for the three axial separations considered, the probability of infection is the least for $L_s=6$ft (\hyperref[fig5]{Figure~\ref{fig5}}).

Note that the above results are based on the “directed jet” behaviour of speech flow (which is the main cause of the sustained aerosol transport away from the speaker’s mouth) generated by the repetition of the phrase “Peter Piper picked a peck” (Abkarian et al., \hyperlink{bib1}{2020}). Abkarian et al. (\hyperlink{bib1}{2020}) also considered other phrases containing fricatives and plosives, and found that the flow associated with the utterance of isolated syllables could be directed upward or downward from the axis at angles as high as $40-50^{\circ}$. However, these flow patterns are contained within a distance of 0.5m from the speaker and die out within 100ms after they are expelled from the mouth (Abkarian et al., \hyperlink{bib1}{2020}). For two people separated by more than 0.6m and for time scales of interaction of the order of 1-2mins as considered here, the transient aerosol transport associated with individual syllables is likely to be less relevant. It will of course be better if the listener can stay entirely outside the zone of influence of the speech flow as suggested by Abkarian et al. (\hyperlink{bib1}{2020}), although it may not be always possible.

Based on our simulations (and taking cues from previous studies), we formulate the following additional guidelines (apart from physical distancing) for short unmasked conversations.
\begin{itemize}
	\item Let the other person speak! Any two-way conversation is far better than a monologue. The one who talks less is at a higher risk of infection. 
	\item In close-up conversations, a slight tilt of the head or a small lateral separation, even while maintaining eye contact, will go a long way in reducing risk.  
	 \item As expected, a greater separation is better, not only while coughing/sneezing but even during short conversations. 
\end{itemize}

The present results can also provide useful inputs to epidemiological models to improve their prediction accuracy. In this connection, Chaudhury et al. (\hyperlink{bib10}{2020})  have presented a framework for determining the infection rates for different expiratory events such as coughing, sneezing, breathing, talking etc. and incorporating them in an SEIRD (susceptible-exposed-infected-recovered-deceased) model (see also Dbouk and Drikakis, \hyperlink{bib17}{2021}). They have highlighted the need for accurate data on viral exposure based on fluid dynamical simulation of these expiratory events. Since the present simulations provide a spatio-temporal distribution of the aerosol field for speech flows, the quantities of interest can be easily calculated, e.g., the temporal variation of the viral load presented in \hyperref[fig7]{Figure~\ref{fig7}} can be used for determining time-dependent infection probabilities for speech flows, which is a crucial input to the model used in Chaudhuri et al. (\hyperlink{bib10}{2020}) . For realistic conversation scenarios these probabilities could not have been accurately inferred from the idealized cases of “one-way” conversation (which would overestimate the risk) or “symmetric” conversation (which would underestimate the risk).

The limitations of the present work should be noted. Parameters like $c_v$ could have different magnitudes than those used here (Wölfel et al., \hyperlink{bib47}{2020}). Also, the number of droplets exhaled is a function of the loudness of speech (Asadi et al., \hyperlink{bib3}{2019}) and can introduce variability in the number of virions exhaled by an infected person. The results presented here are relevant for poorly ventilated spaces whereas a cross ventilation can make a difference. Finally, the speech cycles used are based on utterance of simple repetitive phrases and therefore can only serve as an approximation to the real speech patterns. Notwithstanding these, our simulations provide general scientific principles on which public health guidelines can be based. However, in view of the large number of variables involved, mask mandates are crucial in curbing disease transmission (Chu et al., \hyperlink{bib13}{2020}).

\section{Summary of key results}
We have performed direct numerical simulations of the turbulent transport of aerosols by the flows generated in human speech (involving repetitive utterance of certain phrases) during short conversations. Our main results are: 
\begin{enumerate}
    \item We have computed the total exposure to aerosol on a listener due to speech flow from a speaker not only through inhalation but also through the mouth and eyes. A conservative estimate of the probability of infection is calculated based on the total number of ingested virions and is shown to be higher by a factor of 2 or more compared to previous estimates.
    
    \item In conversations, the active involvement of both people significantly lowers the aerosol exposure (and the associated risk of infection) with respect to the case where one person is a passive listener. This is because of the interference between the two speech jets, which effectively shuts off onward transmission of infected aerosols. On the other hand, asymmetry in the speech pattern of two people, invariably present in real conversations, reduces the jet interference and enhances the risk of infection to the person who talks less.
    
	\item For identical speech phrases from both speakers, the probability of infection peaks when there is a small vertical/lateral separation between them, due to a less effective interaction between the two jets. When this separation is greater than 0.3 times the axial (perpendicular) distance between them, the risk of infection is seen to be minimal.
	
	\item The collision of the speech jets results in a considerable lateral spread of the aerosol cloud potentially infecting surrounding air in an indoor environment. For small axial separations between two people, the speech jets oscillate considerably in time and can penetrate to reach the other person’s face in an intermittent fashion.
\end{enumerate}
	
We believe simulations such as the ones presented here can provide realistic estimates of infection probabilities for any speech pattern of interest and can provide useful inputs to epidemiological models. Lastly, the sensitive dependence of viral exposure during short conversations on several parameters studied here provides insights into the complexity of deciding foolproof public health guidelines to curb the spread of SARS-CoV-2.

\appendix
\begin{appendices}
\section{Governing equations and numerical method}\label{appendix_A}

The Boussinesq Navier-Stokes equations, with scalar transport equations for temperature and passive scalar, are given as 
\begin{equation}
   \vec{\tilde{\nabla}}\  \cdot \  \vec{\tilde {u}} \ =0,
\end{equation}
\begin{equation}
   D{\vec{\tilde {u}}}/D\tilde{t} = -\vec{\tilde{\nabla}} \tilde{p} + \frac{1}{Re} \vec{\tilde{\nabla}}^2 \vec{\tilde {u}} + \frac{1}{Fr^2} \tilde{\theta}
   e_y,
  \end{equation}
\begin{equation}
   D\tilde{\theta} /D\tilde{t} =  \frac{1}{Re \cdot Pr} \tilde{\nabla}^2 \tilde{\theta},
 \end{equation}
\begin{equation}
D\tilde{C_s}/D\tilde{t} =  \frac{1}{Re\cdot Sc} \tilde{\nabla} ^2 \tilde{C_s} ,
\end{equation}
where $\tilde u $ is the fluid velocity,  $\tilde p $ is the pressure, $ \tilde  \theta $ is the temperature difference between the flow and the ambient, $\tilde {C_s} $  is the passive scalar concentration, $D/Dt$ is the material derivative, $Re=u_c d/\nu$ is the Reynolds number, $Fr^2  = u_c^2  / \beta dg \Delta T_o$ is the Froude-number squared, $\nu$ is kinematic viscosity of air, $g$ is acceleration due to gravity, $\beta$ is the coefficient of thermal expansion for air and non-dimensional quantities are represented by a tilde $\textasciitilde $.
The density difference between the speech fluid and the ambient is given by $\Delta\rho⁄\rho =-\beta\Delta T$; with $\beta\approx3.4\times10^{-3}$ 1/K and a temperature difference at the orifice, $\Delta T_o=14^\circ$C, we get $\Delta\rho⁄\rho\approx -0.05$, supporting the Boussinesq approximation. The passive scalar at the orifice is taken as unity, $C_{so}=1$.  The governing parameters are the Prandtl number, $Pr$, and the Schmidt number $Sc$ governing the diffusion of the temperature and scalar respectively, which are both assigned unit values. For the length and velocity scales mentioned above characteristic of respiratory flows, $Re=1906$ and $Fr^{-2}  = 0.00842$. For the conversation cases (with both the persons speaking) two equations are solved for the passive scalar – one each for $\tilde{C_{s1}} $ and $\tilde{ C_{s2}} $.

The DNS solver Megha-5 which is used previously for canonical jet , plumes (Singhal et al., \hyperlink{bib37}{2021}) and cumulus \& mammatus clouds (Ravichandran {\&} Narasimha, \hyperlink{bib34}{2020}; Ravichandran et al., \hyperlink{bib33}{2020}), is employed here. The code discretizes the governing equations in a Cartesian geometry using a second-order scheme in space and uses a second-order accurate Adams-Bashforth scheme for time-stepping. For more details see Singhal et al. (\hyperlink{bib37}{2020}) and Ravichandran et al. (\hyperlink{bib33}{2020}).
\end{appendices}

\vspace*{5pt}

\begin{Backmatter}

\paragraph{Acknowledgements}
The simulations were carried out at the Supercomputer Education and Research Centre at Indian Institute of Science, Bengaluru. 

\paragraph{Funding Statement}
SR gratefully acknowledges support through the Swedish Research Council grant no. 638-2013-9243. SSD acknowledges support from Indian Institute of Science, Bengaluru in the form of the start-up grant no. 1205010620.


\end{Backmatter}

\end{document}